\documentclass[12pt]{article}
\usepackage{amssymb}
\usepackage{epsfig}
\begin{document}
%%%%%%%%%%%%%%%%
%
\setlength{\baselineskip}{0.65cm}
\setlength{\parskip}{0.35cm}
\renewcommand{\thesection}{\Roman{section}}
%
%%%%%%%%%%%%%%%%%
\begin{titlepage}
%%%%%%%%%%%%%%%%%
%
\begin{flushright}
BNL-NT-04/11 \\
RBRC-412 \\
April 2004
\end{flushright}

\vspace*{1.1cm}
\begin{center}
\LARGE

{\bf {Single-Inclusive Jet Production}}\\[2mm]

\medskip
{\bf {in Polarized $\mathbf{pp}$ Collisions at 
$\mathbf{{\cal{O}}(\alpha_s^3)}$ }}\\

\vspace*{2.5cm}
\large 
{B.\ J\"{a}ger$^{a}$, M.\ Stratmann$^{a}$, and W.\ Vogelsang$^{b,c}$}

\vspace*{1.0cm}
\normalsize
{\em $^a$Institut f{\"u}r Theoretische Physik, Universit{\"a}t Regensburg,\\
D-93040 Regensburg, Germany}\\

\vspace*{0.5cm}
{\em $^b$Physics Department, Brookhaven National Laboratory,\\
Upton, New York 11973, U.S.A.}\\

\vspace*{0.5cm}
{\em $^c$RIKEN-BNL Research Center, Bldg. 510a, Brookhaven 
National Laboratory, \\
Upton, New York 11973 -- 5000, U.S.A.}
\end{center}

\vspace*{2.0cm}
%%%%%%%%%%%%%%%%
\begin{abstract}
%%%%%%%%%%%%%%%%                                               
\noindent
We present a next-to-leading order QCD calculation for 
single-inclusive high-$p_T$ jet production in 
longitudinally polarized $pp$ collisions within the ``small-cone''
approximation. The fully analytical expressions obtained for the underlying 
partonic hard-scattering cross sections greatly facilitate the analysis
of upcoming BNL-RHIC data on the double-spin asymmetry
$A_{\mathrm{LL}}^{\mathrm{jet}}$ for this
process in terms of the unknown polarization of gluons in the nucleon.
We simultaneously rederive the corresponding QCD corrections to 
unpolarized scattering and confirm the results existing in the literature. 
We also numerically compare to results obtained with Monte-Carlo methods 
and assess the range of validity of the ``small-cone'' approximation for 
the kinematics relevant at BNL-RHIC.
\end{abstract}
\end{titlepage}
\newpage
%
%%%%%%%%%%%%%%%%%%%%%%
\section{Introduction}
%%%%%%%%%%%%%%%%%%%%%%
%
The first successful runs of the BNL-RHIC with polarized proton 
beams mark a new era in spin physics. 
Very inelastic $pp$ collisions at high energies open up unique
possibilities to answer many interesting questions,
first and foremost perhaps those concerning the still unknown 
gluon polarization in the nucleon, $\Delta g$ \cite{ref:rhic}. 
In the near-term this prime goal can be 
accomplished by studying double spin asymmetries $A_{\mathrm{LL}}$ for 
the inclusive production of hadrons and jets. 
Both are copiously produced at high energies, and 
luminosity requirements are rather modest for a first 
determination of $\Delta g$. Measurements of the unpolarized neutral 
pion cross section at a center-of-mass (c.m.s.) energy of $\sqrt{S}=200\,
\mathrm{GeV}$, for both central and forward pseudo-rapidities
$\eta$ by {\sc{Phenix}} \cite{ref:phenixpion} and {\sc{Star}} 
\cite{ref:starpion}, respectively, have shown good agreement with 
perturbative QCD (pQCD) calculations, even down to unexpectedly small
values of the pion transverse momentum $p_T$ of about 
$1.5\,\mathrm{GeV}$. This boosts confidence that
similar measurements with polarized protons, too, may be
interpreted in terms of pQCD. Very recently, {\sc{Phenix}} has published
first results for $A_{\mathrm{LL}}$ in $pp\rightarrow \pi^0 X$ at 
moderate $p_T$, 
though preliminary and with rather limited statistics \cite{ref:phenixall}.
Surprisingly, the data show a trend towards a negative and sizable
$A_{\mathrm{LL}}$ which, if it persists, would appear to be impossible to 
accommodate in the framework of leading twist pQCD
\cite{ref:ourprl}. More data are clearly needed before any 
definitive conclusions can be drawn. It is expected that
{\sc{Star}} will soon publish data on the spin asymmetry $A_{\mathrm{LL}}^{\mathrm{jet}}$
for the closely related jet production. It will be particularly 
interesting to see whether these data will show a similar trend
as the {\sc{Phenix}} $\pi^0$ data. This paper presents new theoretical pQCD
calculations and predictions for the spin asymmetry in jet
production, which appears timely in view of the experimental
situation. 

In order to make reliable quantitative predictions and to analyze 
upcoming data in terms of spin-dependent parton densities it is 
crucial to know the next-to-leading order (NLO) QCD corrections to 
the lowest order (LO) Born approximation for the partonic
hard-scatterings. NLO corrections can be sizable and may affect 
the spin asymmetries. More importantly, the dependence on unphysical 
factorization and renormalization scales is expected to be
reduced by the inclusion of higher order corrections.
In case of jet production, NLO corrections 
are also of particular importance since it is only at NLO that the 
QCD structure of the jet starts to play a role in the theoretical 
description, providing for the first time the possibility and necessity 
to match the experimental conditions imposed to group final-state
partons into a jet.

The past few years have seen much progress in calculations of 
NLO QCD corrections to spin-dependent processes relevant for the 
RHIC spin program. In particular for inclusive hadron production, 
NLO results have been obtained recently both at an analytical level 
\cite{ref:nlopion} as well as using Monte-Carlo integration 
techniques \cite{ref:daniel}.  While the latter basically allow to 
compute any infrared-safe cross section numerically, the former is 
restricted to inclusive hadron spectra. However, fully analytical 
calculations lead to much faster and more
efficient computer codes as all singularities arising in intermediate steps 
have explicitly canceled and are not subject to delicate numerical treatments.
Such calculations will greatly facilitate a future ``global analysis'' of RHIC
data in terms of polarized parton densities, e.g., along the lines described
in \cite{ref:global}.

The aim of this paper is to obtain (approximate) 
{\em analytical} NLO QCD results also in the case of 
single-inclusive high-$p_T$ jet production in longitudinally 
polarized $pp$ collisions. NLO corrections to polarized jet production have 
been determined before, using Monte-Carlo methods \cite{ref:mcjets}. 
However, for the reason just given, we believe that our calculation is 
a very useful addition and of great relevance in practice. Its results
will be immediately usable in a phenomenological analysis of 
forthcoming {\sc{Star}} data.

The actual calculation of the partonic cross sections up to NLO is 
a formidable task. Fortunately, inclusive hadron and jet production 
proceed through the same set of partonic subprocesses, and it turns
out that we can make use of much of what we already calculated 
previously in the case of hadron production \cite{ref:nlopion}. 
The main difference is in the 
treatment of final-state singularities: in case of inclusive hadron 
production the phase space of all 
unobserved partons is fully integrated over. This leads to final-state 
singularities which are to be absorbed into the parton-to-hadron 
fragmentation functions. However, a jet is essentially defined as
a transverse-energy deposition within a certain cone centered 
around pseudo-rapidity $\eta^{\mathrm{jet}}$ and azimuth 
$\phi^{\mathrm{jet}}$. For such an observable, all final-state singularities 
cancel \cite{ref:sterwbg}. As we will see below, despite this difference, 
it is possible to transform the single-parton-inclusive cross sections 
that we have calculated earlier \cite{ref:nlopion} into single-inclusive 
jet cross sections. This can even be done at an analytical level if one 
assumes that the jet cone is rather narrow (``small-cone 
approximation'', SCA). This is the strategy we will use in this paper. 
We note that the SCA was first introduced many years 
ago \cite{ref:sterwbg,ref:furman} and has been applied in computations of 
unpolarized single-inclusive jet cross sections \cite{ref:aversa}. 
We have re-done the calculations in the unpolarized case 
and fully confirm the results available in the literature \cite{ref:aversa}. 
By comparing our polarized cross sections obtained within the SCA to results
of the NLO Monte-Carlo ``parton generator'' of \cite{ref:mcjets} 
we can assess the range of applicability of the SCA. It turns out that
for the kinematics relevant at RHIC the SCA is very accurate,
and that our numerical code is orders of magnitude faster than 
the Monte-Carlo one. Thus our results will indeed be useful for the 
analysis of upcoming data in terms of polarized parton densities.

The outline of the paper is as follows: after defining our notation, 
the next section is mainly devoted to present in some detail the
necessary technical framework to convert single-parton-inclusive cross sections
into jet cross sections. In Sec.\ III we present some phenomenological 
applications of our results, all tailored to upcoming measurements with
the {\sc Star} experiment at the BNL-RHIC. First we compare the SCA to 
the Monte-Carlo results obtained for various cone sizes. Next 
we examine the size of the NLO corrections, the reduction of dependence on 
unphysical scales in NLO, and the sensitivity of the double-spin 
asymmetry $A_{\mathrm{LL}}$ to $\Delta g$. 
We summarize the main results in Sec.\ IV.

%%%%%%%%%%%%%%%%%%%%%%%%%%%%%%%%%%%%%%%%%%%%%%%%%%%%%%%%
\section{Jet Production in the Small-Cone Approximation}
%%%%%%%%%%%%%%%%%%%%%%%%%%%%%%%%%%%%%%%%%%%%%%%%%%%%%%%%
%
%%%%%%%%%%%%%%%%%%%%%%%%%%%%%%%%%%%%%%%%%%%%%%%%%%%%
\subsection{Jet Definition and SCA}
\begin{figure}[th]
\begin{center}
\epsfig{figure=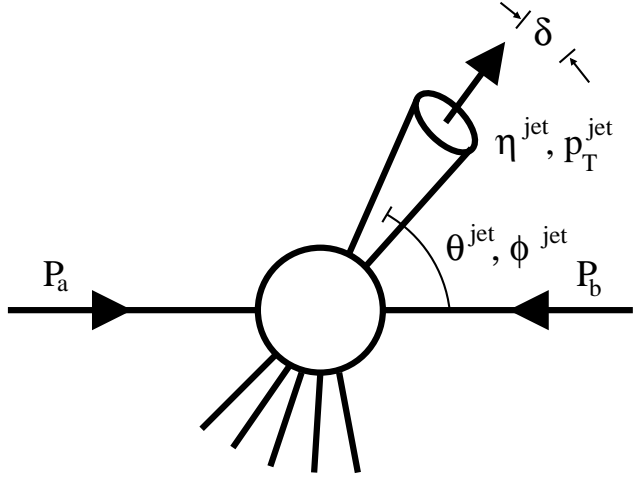,angle=0,width=0.50\textwidth}
\end{center}
\vspace*{-0.3cm}
\caption{\label{fig:fig1} \sf Sketch of single-inclusive jet production in
$p(P_a) p(P_b) \rightarrow \mathrm{jet}(P_{\mathrm{jet}})\,X$.}
\end{figure}

%%%%%%%%%%%%%%%%%%%%%%%%%%%%%%%%%%%%%%%%%%%%%%%%%%%%
%
We will consider the single-inclusive jet cross section for the
reaction $pp\rightarrow \mathrm{jet}\,X$ which counts all
events with a jet of a given transverse energy
$E_T^{\mathrm{jet}}$ and pseudo-rapidity $\eta^{\mathrm{jet}}$.
Cross sections are infinite unless a finite jet ``size'' is imposed.
Jets are not intrinsically well defined objects, but may be
constructed in somewhat different fashions from a set of close-by 
final-state particles. In this work we define the jet four-momentum
as the sum of the four-momenta of all particles inside a 
geometrical cone of half-aperture $\delta$ around the jet axis,
given by the three-momentum of the jet, see Fig.~\ref{fig:fig1}. 
We treat the jet in the SCA \cite{ref:furman,ref:aversa}
as this allows for a largely analytical calculation of the cross section.

Other jet definitions and algorithms are clearly possible and, in fact,
more widely used in practice [13-17].
%\cite{ref:eks,ref:snowmass,ref:othercone,ref:cluster,ref:discussion}.
A popular choice \cite{ref:snowmass}, also adopted in the jet 
analyses at STAR, is to define a jet as a deposition of total transverse 
energy $\sum_i E_T^i$ of all final-state particles that fulfill
\begin{equation}
\label{eq:conedef}
(\eta^{\mathrm{jet}}-\eta^i)^2 + (\phi^{\mathrm{jet}}-\phi^i)^2 \le R^2 \; .
\end{equation}
Here $\eta^i$ and $\phi^i$ denote the pseudo-rapidities and azimuthal
angles of the particles, and $R$ is the jet cone aperture. 
The jet variables are defined by $E_T^{\mathrm{jet}}=\sum_i E_T^i$, 
$\eta^{\mathrm{jet}}=-\ln[\tan (\theta^{\mathrm{jet}}/2)]=
\sum_i E_T^i \eta^i / E_T^{\mathrm{jet}}$, and
$\phi^{\mathrm{jet}}=\sum_i E_T^i \phi^i / E_T^{\mathrm{jet}}$. 
It has been shown \cite{ref:salesch} that in the SCA the jet definition
we have adopted becomes equivalent to that given by Eq.~(\ref{eq:conedef}),
provided $\delta$ is chosen as $R/ \cosh\left(\eta^{\mathrm{jet}}\right)$.
We have verified this equivalence numerically using the Monte-Carlo
code of \cite{ref:mcjets}. Differences arise only at ${\cal{O}}(\delta^2)$ 
and are negligible for small cone apertures. 

%%%%%%%%%%%%
% FIGURE 1 %
%%%%%%%%%%%%
The SCA may be viewed as an expansion of the partonic cross sections
around $\delta =0$ (or, equivalently, $R=0$). For small $\delta$, the 
dependence on the cone size is of the form ${\cal A}\log(\delta)+{\cal B}+{\cal O}
(\delta^2)$. In this work we determine the coefficients ${\cal A}$ 
and ${\cal B}$ at an analytical level. We neglect the ${\cal O}(\delta^2)$ pieces; 
we will demonstrate in Sec.~III that this is a surprisingly good approximation even 
for (experimentally relevant) cone sizes of up to $R\simeq 0.7$. This observation was 
also made some time ago for unpolarized jet cross sections by comparing the SCA to a 
calculation where the SCA was extended numerically to finite cone sizes 
\cite{ref:scavalid,ref:guillet}.

%
%%%%%%%%%%%%%%%%%%%%%%%%%%%%%%%%%%%%%%%%%%%%%%%%%%%%
\subsection{Notation and Outline of the Calculation}
%%%%%%%%%%%%%%%%%%%%%%%%%%%%%%%%%%%%%%%%%%%%%%%%%%%%
%
According to the factorization theorem for high-$p_T$ processes \cite{ref:fact}, 
the spin-dependent cross section for single-inclusive jet production
$p(P_a)p(P_b)\rightarrow\mathrm{jet}(P_{\mathrm{jet}})\,X$ can be written as
\begin{eqnarray}
\label{eq:xsecdef}
\frac{d\Delta\sigma}{dp_T^{\mathrm{jet}}d\eta^{\mathrm{jet}}} &=& \frac{1}{2} \left[ 
\frac{d\sigma^{++}}{dp_T^{\mathrm{jet}} d\eta^{\mathrm{jet}}} - 
\frac{d\sigma^{+-}}{dp_T^{\mathrm{jet}} d\eta^{\mathrm{jet}}} \right] \\[3mm]
&=& \frac{2p_T^{\mathrm{jet}}}{S} \sum_{a,b} 
\int_{VW}^{V}\frac{dv}{v(1-v)} 
\int^1_{VW/v}\!\!\frac{dw}{w}\Delta f_a(x_a,\mu_F) \Delta f_b(x_b,\mu_F) 
\nonumber \\[3mm]
&& \times \left[ 
\frac{d\Delta \hat{\sigma}^{(0)}_{ab\rightarrow \mathrm{jet} X}(s,v)}{dv} 
\delta (1-w) + \frac{\alpha_s(\mu_R)}{\pi} \, \frac{d\Delta 
\hat{\sigma}^{(1)}_{ab\rightarrow \mathrm{jet}X}(s,v,w,\mu_F,\mu_R; \delta)}
{dvdw}
\right] \label{eq:xsecfact} \;, 
\end{eqnarray}
where the superscripts in Eq.~(\ref{eq:xsecdef}) denote the helicities
of the colliding polarized protons. The sum in Eq.~(\ref{eq:xsecfact}) 
is over all contributing partonic channels $a+b\to \mathrm{jet}+X$,
with their associated LO and NLO partonic cross sections 
$d\Delta\hat{\sigma}^{(0)}_{ab\rightarrow \mathrm{jet} X}$ and 
$d\Delta\hat{\sigma}^{(1)}_{ab\rightarrow \mathrm{jet} X}$, respectively.
The latter are defined in complete analogy with Eq.~(\ref{eq:xsecdef}),
the helicities now referring to partonic ones. In Eq.~(\ref{eq:xsecfact}) 
we have introduced the dimensionless variables $V$ and $W$ which are 
defined in terms of $p_T^{\mathrm{jet}}$ and $\eta^{\mathrm{jet}}$ as
\begin{equation}
V=1-\frac{p_T^{\mathrm{jet}}}{\sqrt{S}}\; e^{\eta^{\mathrm{jet}}}\;\;\;\mathrm{and}\;\;\;
W=\frac{{(p_T^{\mathrm{jet}})}^2}{S V (1-V)}\;\;\;,
\label{eq:vwdef}
\end{equation}
with $S=(P_a+P_b)^2$ the available hadronic c.m.s.\ energy squared.
The corresponding parton-level variables read
\begin{equation}
v\equiv 1+\frac{t}{s}\;\;,\;\; w\equiv \frac{-u}{s+t}\;\;,\;\;
s\equiv (p_a+p_b)^2\;\;,\;\;t\equiv (p_a-P_{\mathrm{jet}})^2\;\;,\;\;u\equiv
(p_b-P_{\mathrm{jet}})^2\;\;,
\label{eq:partvar}
\end{equation}
where $P_{\mathrm{jet}}$ is the four-momentum of the jet 
($P_{\mathrm{jet}}^2\simeq 0$ in the SCA). The information 
on the spin structure of the proton in Eq.~(\ref{eq:xsecfact})
is contained in the spin-dependent parton densities $\Delta f_{a,b}$.
They are probed at momentum fractions $x_{a,b}$ given by
\begin{equation}
\label{eq:xaxb}
x_a = \frac{V W}{v w}\;\;\;\mathrm{and}\;\;\;
x_b=  \frac{1-V}{1-v}\;\;\;.
\end{equation}
The factorized structure of Eq.~(\ref{eq:xsecfact}) dictates the
appearance of the factorization scale $\mu_F$ which is of the order
of the hard scale in the reaction, $p_T^{\mathrm{jet}}$, but
not further specified. The same is true for the scale $\mu_R$ associated with
the renormalization of the running strong coupling $\alpha_s(\mu_R)$.
Compared to single-inclusive hadron production \cite{ref:nlopion,ref:daniel}
the jet observables in Eq.~(\ref{eq:xsecfact}) have the advantage of being 
free of uncertainties associated with the factorization of final-state 
singularities into non-perturbative parton-to-hadron fragmentation 
functions at a scale $\mu^{\prime}_F$. Hence, jet production at RHIC
is expected to provide particularly clean and useful information on 
the spin structure of the proton. 

As indicated above, the jet cross sections at the parton-level may be
evaluated in pQCD as an expansion in the strong coupling
$\alpha_s$,
\begin{equation}
\label{eq:partxsec}
d\Delta\hat{\sigma}_{ab\rightarrow \mathrm{jet}X}=
d\Delta\hat{\sigma}_{ab\rightarrow \mathrm{jet}X}^{(0)} + \frac{\alpha_s}{\pi}
d\Delta\hat{\sigma}_{ab\rightarrow \mathrm{jet}X}^{(1)} + \;\ldots\;\;\;.
\end{equation}
The LO approximation of Eq.~(\ref{eq:partxsec}), 
$d\Delta\hat{\sigma}_{ab\rightarrow \mathrm{jet}X}^{(0)}$, is obtained from 
evaluating all basic $2\rightarrow 2$ QCD scattering diagrams, that is, $X$
consists of only one parton recoiling from the other parton producing the
observed jet. Contrary to single-inclusive hadron production where different
final-state partons have to be weighted with different fragmentation functions
leading to ten separate LO channels \cite{ref:nlopion}, for a jet cross 
section no distinction is made between different quark flavors or gluons
producing the jet. All processes with the same initial-state partons have 
to be summed appropriately, which is precisely the reason why 
final-state singularities cancel at higher orders. One then 
has only six different subprocesses~:
%
%%%%%%%%%%%%
% FIGURE 2 %
%%%%%%%%%%%%
\begin{figure}[th]
\begin{center}
\begin{minipage}{8cm}
\begin{center}
\epsfig{figure=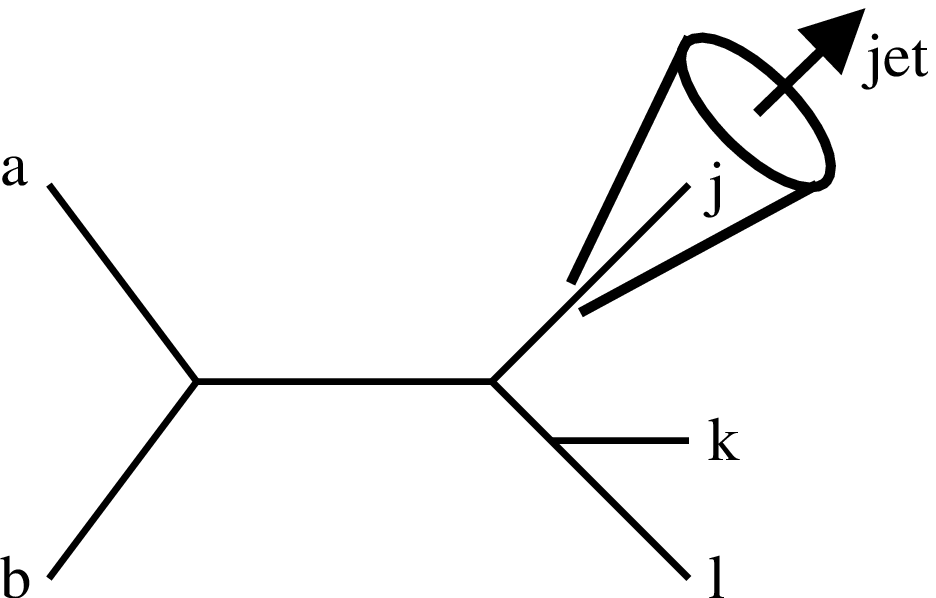,angle=0,width=0.80\textwidth}
\end{center}
\end{minipage}
\begin{minipage}{8cm}
\begin{center}
\epsfig{figure=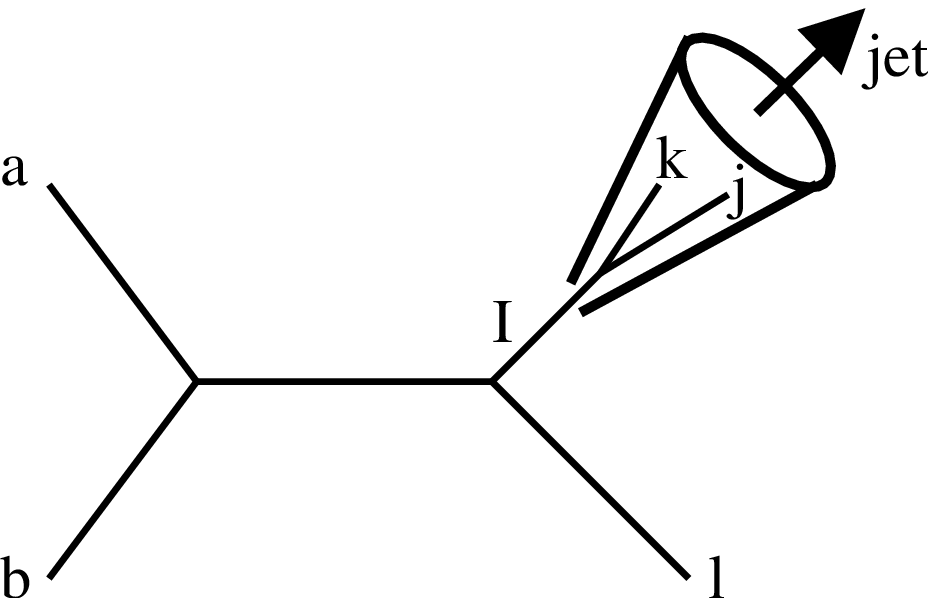,angle=0,width=0.80\textwidth}
\end{center}
\end{minipage}
\end{center}
\caption{\sf Contributions to the single-inclusive jet cross section from
partonic reactions where only one parton is in the cone (left hand side)
and where two essentially collinear partons, $j$ and $k$, form
a narrow jet (right hand side).
\label{fig:fig2}}
\end{figure}
\begin{eqnarray}
qq'             &\to& \mathrm{jet} \, X\nonumber\\
q\bar{q}'       &\to& \mathrm{jet} \, X\nonumber\\
qq              &\to& \mathrm{jet} \, X\nonumber\\
q\bar{q}        &\to& \mathrm{jet} \, X\nonumber\\
qg              &\to& \mathrm{jet} \, X\nonumber\\
gg              &\to& \mathrm{jet} \, X\;\;.
\label{eq:proclist}
\end{eqnarray}

For the computation of the NLO corrections to Eq.~(\ref{eq:partxsec}), 
$d\Delta\hat{\sigma}_{ab\rightarrow \mathrm{jet}X}^{(1)}$, we would like to
make use as much as possible of the single-parton-inclusive cross sections 
that we have calculated earlier \cite{ref:nlopion}. 
In the following we will demonstrate how to convert {\em analytically} the
single-parton-inclusive cross sections into jet cross sections.
In order to achieve this, we use the SCA and that
the jet is formed either by a single final-state parton or by two
partons essentially collinear to each other, see Fig.~\ref{fig:fig2}.

The connection between the two types of inclusive partonic cross sections 
can be best understood in the following way \cite{ref:furman,ref:guillet}: 
let us consider the generic
$2\rightarrow 3$ partonic reaction $a(p_a)\,b(p_b)\rightarrow
j(p_j)\,k(p_k)\,l(p_l)$. Partons $a$ and $b$ are longitudinally polarized
here, which is however not really of any relevance in what follows.  
In \cite{ref:nlopion} we have already computed at an analytical level the
single-parton inclusive cross sections for $ab\to jX$, $ab\to kX$, 
and $ab\to lX$. Each one of these already contains a full sum over 
unobserved partonic final states $X$, that is, $ab\to jX$ would 
consist of $[ab\to j (k_1 l_1)] + [ab\to j (k_2 l_2)] + \ldots$, in case
several different final states are allowed. For instance, the channel 
$qg\to qX$ receives contributions from the partonic
reactions $qg\to q (gg)$, $qg\to q(q\bar{q})$, $qg\to q(q'\bar{q}')$.
With these stipulations, all three inclusive-parton processes, 
$ab\to jX$, $ab\to kX$, and $ab\to lX$, contribute also to 
single-inclusive jet production, for which they have to be added. 

Looking closer at the cross section for $ab\to jX$ one has to distinguish
configurations where only parton $j$ is in the cone and forms the
jet from situations where, for example, partons $j$ and
$k$ (and similarly, $j$ and $l$) are both in the cone, 
see Fig.~\ref{fig:fig2}. In the latter case, the single-parton-inclusive 
cross section still assumes that only parton $j$ is observed, whereas 
for the jet cross section we want to define the jet from {\em both} 
partons inside the cone. We therefore subtract those parts from the 
single-parton-inclusive cross section for which $j$ is observed, 
but $k$ (or $l$) are in the cone. We proceed, of course, in the 
same way for the cross sections with $k$ or $l$ being the observed 
parton. Finally, we have to {\em add} pieces for which $j$ and $k$ 
(or $j$ and $l$, or $k$ and $l$) are both in the cone, and for 
which $j$ and $k$ are forming the jet together. Note that these
contributions are symmetric under exchange of $j$ and $k$.

To be more specific, let us denote the longitudinally polarized
single-parton-inclusive cross section for $ab\rightarrow jX$ by 
$d\Delta \hat{\sigma}_j$, the one where still $j$ is observed, 
but $k$ is also in the cone by $d\Delta \hat{\sigma}_{j(k)}$, and the one 
where $j$ and $k$ are in the cone and form the jet {\em together} by 
$d\Delta \hat{\sigma}_{jk}$. We then have the following final expression 
for the desired partonic jet cross section:
\begin{eqnarray}
\nonumber
d\Delta\hat{\sigma}_{ab\rightarrow \mathrm{jet}X} &=&  
[d\Delta\hat{\sigma}_j -d\Delta\hat{\sigma}_{j(k)}-
d\Delta\hat{\sigma}_{j(l)}]
+ [d\Delta\hat{\sigma}_k -d\Delta\hat{\sigma}_{k(j)}-
d\Delta\hat{\sigma}_{k(l)}]\\
&& + [d\Delta \hat{\sigma}_l -d\Delta\hat{\sigma}_{l(j)}-
d\Delta\hat{\sigma}_{l(k)}]+
d\Delta\hat{\sigma}_{jk} + d\Delta\hat{\sigma}_{jl}+ 
d\Delta\hat{\sigma}_{kl} \; .
\label{eq:jetoutline}
\end{eqnarray}
All final-state singularities of the individual cross sections 
are guaranteed to cancel in Eq.~(\ref{eq:jetoutline}).
Technically, because the $d\Delta\hat{\sigma}_j$ have already
been made finite by an $\overline{\rm{MS}}$ subtraction \cite{ref:nlopion},
we will first compute the combination
\begin{equation} 
\label{eq:subcombi}
-d\Delta\hat{\sigma}_{j(k)}-d\Delta\hat{\sigma}_{j(l)}
-d\Delta\hat{\sigma}_{k(j)}-d\Delta\hat{\sigma}_{k(l)}
-d\Delta\hat{\sigma}_{l(j)}-d\Delta\hat{\sigma}_{l(k)}+
d\Delta\hat{\sigma}_{jk} + d\Delta\hat{\sigma}_{jl}+ 
d\Delta\hat{\sigma}_{kl} \;,
\end{equation}
which will have final-state collinear singularities,
the same (but with opposite sign) that were originally present in 
$d\Delta\hat{\sigma}_j + d\Delta\hat{\sigma}_k + d\Delta\hat{\sigma}_l$.
To obtain a finite answer for the combination (\ref{eq:subcombi})
an $\overline{\rm{MS}}$ subtraction then has to be performed.
A powerful check for the correct implementation of all contributions
is to see that the dependence on the final-state factorization scale
$\mu^\prime_F$ present in the intermediate results
disappears identically for each of the subprocesses listed in Eq.~(\ref{eq:proclist})
when $d\Delta\hat{\sigma}_j + d\Delta\hat{\sigma}_k + d\Delta\hat{\sigma}_l$, 
taken from \cite{ref:nlopion}, is finally
combined with (\ref{eq:subcombi}). 

Clearly, the pieces $d\Delta \hat{\sigma}_{j(k)}$ and 
$d\Delta \hat{\sigma}_{jk}$ that we will subtract and add are dominated by 
configurations with two collinear final-state partons. These configurations
produce the ${\cal A}\log(\delta)+{\cal B}+{\cal O}
(\delta^2)$ behavior in $\delta$ mentioned earlier. 
For collinear kinematics, the calculation 
dramatically simplifies because the $2\to 3$ matrix elements then 
factorize into $2\to 2$ ones and LO splitting functions, allowing
the calculation to be done largely analytically. The narrower 
one chooses the cone, the more dominant will the ``collinear'' terms 
become. This is the reasoning behind the use of the SCA.

We note that the separation in Eq.~(\ref{eq:jetoutline}) in the 
context of the SCA was already considered in 
Refs.~\cite{ref:furman,ref:guillet} in the unpolarized case. 
In these references also some details 
of the calculations of the $d\hat{\sigma}_{j(k)}$ and 
$d \hat{\sigma}_{jk}$ were given. The aim of the following 
subsections is to make this paper self-contained by providing all
essential calculational details, without however going to 
excessive length. In the next two subsections, we will address
$d\Delta \hat{\sigma}_{j(k)}$ and $d\Delta \hat{\sigma}_{jk}$
separately.

%%%%%%%%%%%%%%%%%%%%%%%%%%%%%%%%%%%%%%%%%%%%%%%%%%%%%%%%%%%%%%%%%
\subsection{Calculation of $\mathbf{d\Delta\hat{\sigma}_{j(k)}}$}
%%%%%%%%%%%%%%%%%%%%%%%%%%%%%%%%%%%%%%%%%%%%%%%%%%%%%%%%%%%%%%%%%
%
In this section we show how to calculate the ``subtraction'' contributions
$d\Delta\hat{\sigma}_{j(k)}$ to Eq.~(\ref{eq:jetoutline}) when 
parton $j$ alone forms the jet but parton $k$ is also in the 
cone. We note that this part of the calculation is very similar 
to what was done for the isolated prompt photon cross section 
\cite{ref:photcone}, except for the fact that all possible 
parton-parton splittings will occur in our case. The leading 
contributions in the SCA result from a branching of an
intermediate parton $I$ (which could, of course, be of the same
type as $j$ and/or $k$), see Fig.~\ref{fig:fig2}.
In the {\em collinear approximation}, the spin-dependent
matrix element squared $\Delta |M|_{ab\to jkl}^2$ for the process 
$ab\rightarrow jkl$ factorizes into 
\begin{equation} 
\label{eq:collapprox}
\Delta |M|_{ab\to jkl}^2 \propto \frac{1}{2p_j\cdot p_k}\,
\Delta |M|_{ab\to Il}^2 \left(s,v'=\frac{vw}{1-v+vw}\right)
P_{jI}^{<} (z=1-v+vw) \;\;,
\end{equation}
where all quantities are in $d=4-2\varepsilon$ dimensions in order to 
regularize collinear singularities. The superscript ``$<$'' 
denotes that the $d$-dimensional $I\rightarrow j$ 
unpolarized splitting function $P_{jI}(z)$ is 
strictly at $z<1$, that is, without its $\delta(1-z)$ contribution if $I=j$. 
$\Delta |M|_{ab\to Il}^2$ denotes the spin-dependent
matrix element squared for the $2\rightarrow 2$ process $ab\rightarrow I l$.
In addition, we find
\begin{equation}
2p_j\cdot p_k=2 E_j^2 \frac{v\,(1-w)}{1-v+vw} (1-\cos\theta_{jk})\;\;,
\end{equation}
with $\theta_{jk}$ the angle between partons $j$ and $k$.
The factor $1/(2p_j\cdot p_k)$ in (\ref{eq:collapprox}) constitutes the only 
dependence on $\theta_{jk}$ in the collinear approximation.
Integrating it over the $2\to 3$ phase space gives
\begin{eqnarray}
\int \frac{dPS_3}{dvdw}\; \frac{1}{(2p_j\cdot p_k)} &=&
\left[ \frac{1}{8\pi} \left( \frac{4\pi}{s}
\right)^{\varepsilon}\frac{1}{\Gamma(1-\varepsilon)}[v'(1-v')]^{-\varepsilon}
\right]\frac{1}{8\pi^2}\left( \frac{4\pi}{s}
\right)^{\varepsilon} \nonumber \\[3mm]
&&\times\;\frac{1}{\Gamma(1-\varepsilon)}
\frac{v}{1-v+vw}
\left[ \frac{E_j^2 v^2(1-w)^2}{s}\right]^{-\varepsilon} \!
\int_0^{\delta} d\theta_{jk}\frac{\sin^{1-2 \varepsilon}
\theta_{jk}}{1-\cos\theta_{jk}} \;,
\label{eq:sca23ps}
\end{eqnarray}
where $\delta$ is the cone opening, $v'$ is as in Eq.~(\ref{eq:collapprox}), 
and the factor in square brackets in the first line is 
the usual $2\to 2$ phase space in $d$ dimensions. For small $\delta$ 
the angular integral in Eq.~(\ref{eq:sca23ps}) is readily evaluated as
\begin{equation}
\int_0^{\delta} \, d\theta_{jk}\frac{\sin^{1-2 \varepsilon}
\theta_{jk}}{1-\cos\theta_{jk}}=
-\frac{\delta^{-2 \varepsilon}}{\varepsilon}+ {\cal{O}}(\delta^2) \;\;.
\end{equation}
For the $d$-dimensional partonic cross section at 
$\mathcal{O}(\alpha_s^3)$ in the collinear approximation we then arrive at
\begin{eqnarray}
\label{eq:sca23xsec}
\frac{d\Delta\hat{\sigma}_{ab\to jkl}}{dvdw} &=& 
\frac{d\Delta\hat{\sigma}_{ab\to Il}}{dv}
\left( v'=\frac{vw}{1-v+vw}\right)\; P_{jI}^{<} (z=1-v+vw) \nonumber \\[3mm]
&&\times
\frac{\alpha_s}{2\pi}\left(-\frac{1}{\varepsilon}\right)\frac{1}{
\Gamma(1-\varepsilon)}\frac{v}{1-v+vw}\left[
\frac{E_j^2 \delta^2 v^2(1-w)^2}{s}\right]^{-\varepsilon}\; .
\end{eqnarray}
Here, $d\Delta\hat{\sigma}_{ab\to Il}/dv$ denotes the $d$-dimensional
Born cross section for 
$a b\to I l$.  One can observe in Eq.~(\ref{eq:sca23xsec}) the expected 
separation into a universal part associated with the splitting function 
$P_{jI}$, and a part related to an underlying $2\to 2$ partonic scattering.

From now on we will mainly focus on the splitting part, keeping in mind, 
however, that the appropriate sum over all partonic channels and 
splittings is implicitly understood in the end.
The further evaluation is particularly simple when the splitting happens to be
non-diagonal, i.e.\ $I\neq j$.
In that case, the full splitting function $P_{jI}$ is regular at $z=1$, i.e.,
at $w=1$, and all terms may be safely expanded in $\varepsilon$. 
The collinear singularity arising in Eq.~(\ref{eq:sca23xsec}) was also present 
in the single-parton-inclusive cross sections $d\Delta \hat{\sigma}_j$, 
where it was subtracted at a scale $\mu^{\prime}_F$ into the bare 
parton-to-hadron fragmentation functions in the $\overline{\rm{MS}}$ 
scheme \cite{ref:nlopion}. This factorization is not appropriate for
a jet cross section. In order to compensate for it,
we have to apply the same subtraction also to Eq.~(\ref{eq:sca23xsec}).
Any dependence on the factorization scale $\mu^{\prime}_F$ will then
drop out for the inclusive jet cross section in the sum 
(\ref{eq:jetoutline}) over partonic scatterings, as discussed above. 
The appropriate factorization term for Eq.~(\ref{eq:sca23xsec}) reads
\begin{eqnarray} 
\label{eq:scafact}
\nonumber
\frac{d\Delta\hat{\sigma}_{\mathrm{fact}}}{dvdw} &=&
-\frac{\alpha_s}{2\pi}\, \frac{d\Delta\hat{\sigma}_{ab\to Il}}{dv}
\left( v'=\frac{vw}{1-v+vw}\right)\; P^{(4)}_{jI} (z=1-v+vw)\\[2mm]
&&\times 
\left(-\frac{1}{\varepsilon}\right)\frac{v}{1-v+vw}\left( 
\frac{{\mu^{\prime}_F}^2}{s}\right)^{-\varepsilon}\; ,
\end{eqnarray}
where now $P_{jI}^{(4)}$ is the usual four-dimensional splitting
function. Expressing the $d$-dimensional one as
\begin{equation}
P_{jI}(z)=P^{(4)}_{jI}(z)+\varepsilon P^{(\varepsilon)}_{jI}(z) \;\;,
\end{equation}
and adding Eqs.~(\ref{eq:sca23xsec}) and (\ref{eq:scafact}) we arrive
at the final result for $d\Delta\hat{\sigma}_{j(k)}$ if $j\neq I$,
\begin{eqnarray}  
\label{eq:scanondiag}
\frac{d\Delta\hat{\sigma}_{j(k)}}{dvdw} &=& \frac{\alpha_s}{2\pi}\,
\frac{v}{1-v+vw}\, \frac{d\Delta \hat{\sigma}^{(d=4)}_{ab\to Il}}{dv}
\left( v'=\frac{vw}{1-v+vw}\right)\; \nonumber \\[3mm]
&&\times
\left[ P^{(4)}_{jI} (z=1-v+vw) \ln\left( \frac{E_j^2 \delta^2 
v^2 (1-w)^2}{{\mu^{\prime}_F}^2} \right) - 
P^{(\varepsilon)}_{jI}(z=1-v+vw)\right],
\end{eqnarray} 
which is finite and shows the expected logarithmic dependence on 
$\delta$.

For a diagonal splitting function in Eq.~(\ref{eq:sca23xsec}), i.e., $j=I$,
the situation is somewhat more complicated due to the infrared singularity 
of $P_{jj}(z)$ at $z=1$. The singularity is regularized by the factor 
$(1-w)^{-2 \varepsilon}$ in (\ref{eq:sca23xsec}) and gives rise to 
a $1/\varepsilon^2$ pole. This pole is then canceled by contributions 
from $d\Delta\hat{\sigma}_{jk}$, i.e., the cross section with both 
partons in the cone forming the jet. Rather than giving the lengthy
(regularized) expression for the diagonal case, we therefore first turn 
to the calculation of $d\Delta\hat{\sigma}_{jk}$.

%%%%%%%%%%%%%%%%%%%%%%%%%%%%%%%%%%%%%%%%%%%%%%%%%%%%%%%%%%%%%%%
\subsection{Calculation of $\mathbf{d\Delta\hat{\sigma}_{jk}}$}
%%%%%%%%%%%%%%%%%%%%%%%%%%%%%%%%%%%%%%%%%%%%%%%%%%%%%%%%%%%%%%%
%
Again, the leading configurations are those for which an outgoing
intermediate parton $I$ splits collinearly into two. We now
have to deal with partons $j$ and $k$ in the cone
producing the jet with four-momentum $P_{\mathrm{jet}}=p_j+p_k$. 
Let us first note that for the $d$-dimensional $2\rightarrow 3$
phase space one finds
\begin{eqnarray}
2E_{\mathrm{jet}}\frac{dPS_3}{d^{d-1}P_{\mathrm{jet}}}&=&
\frac{1}{(2 \pi)^{5-4 \varepsilon}}\int
d^d p_j \delta(p_j^2) \frac{E_{\mathrm{jet}}}{E_k}\,\delta\!
\left( [p_a+p_b-P_{\mathrm{jet}}]^2 \right) \nonumber \\
&=&\frac{1}{(2 \pi)^{5-4 \varepsilon}}\frac{1}{sv}\delta(1-w)
\int d^d p_j \delta(p_j^2) \frac{E_{\mathrm{jet}}}{E_k}\, \; .
\label{eq:scaps2}
\end{eqnarray}
The cross sections $d\Delta\hat{\sigma}_{jk}$ thus
contribute only at $w=1$, i.e., with $2\rightarrow 2$ kinematics.
As discussed in the previous subsection the only dependence on 
$\theta_{jk}$, the angle between the two partons $j$ and $k$, stems 
from the propagator $\propto 1/(2 p_j\cdot p_k)$. Integrating
this term over the phase space (\ref{eq:scaps2}) one finds after some
algebra:
\begin{eqnarray} 
\label{eq:scaps3}
\int \frac{dPS_3}{dvdw} \frac{1}{(2p_j\cdot p_k)} &=&
\left[ \frac{1}{8\pi} \left( \frac{4\pi}{s}
\right)^{\varepsilon}\frac{1}{\Gamma(1-\varepsilon)}[v(1-v)]^{-\varepsilon}
\right]\frac{1}{8\pi^2}\left( \frac{4\pi}{s}
\right)^{\varepsilon} \delta(1-w)\nonumber \\[3mm]
&&\times\frac{1}{\Gamma(1-\varepsilon)}
\int_0^{E_{\mathrm{jet}}} dE_j \,\frac{E_{\mathrm{jet}}}{E_k^2} 
\left( \frac{E_j^2}{s}\right)^{-\varepsilon}
\int_0^{\theta_{\mathrm{max}}} d\theta_{j}\frac{\sin^{1-2 \varepsilon}
\theta_{j}}{1-\cos\theta_{jk}} \;\;,
\end{eqnarray}
where $\theta_j$ is the angle of parton $j$ with respect to 
the direction of the jet, $\vec{P}_{\mathrm{jet}}$.

In order to find the relation between $\theta_{j}$ and $\theta_{jk}$ 
we first write
\begin{equation}
\cos \theta_{jk} =\frac{\vec{p}_j\cdot\vec{p}_k}{|\vec{p}_j|\,|\vec{p}_k|}=
\frac{\vec{p}_j\cdot\left(\vec{P}_{\mathrm{jet}}-\vec{p}_j\right)}
{E_j\, (E_{\mathrm{jet}}-E_j)}=\frac{|\vec{P}_{\mathrm{jet}}|
\cos\theta_j-E_j}{E_{\mathrm{jet}}-E_j} 
\label{eq:theta12}
\end{equation}
and then use 
\begin{equation}
\label{eq:theta1}
\vec{P}_{\mathrm{jet}}^{2} = (\vec{p}_j+\vec{p}_k)^2 = E_j^2 + 
(E_{\mathrm{jet}}-E_j)^2+2 E_j (E_{\mathrm{jet}}-E_j) \cos\theta_{jk}\;\;.
\end{equation}
Solving Eqs.~(\ref{eq:theta12}) and (\ref{eq:theta1}) for 
$\theta_{j}$, one finds in the collinear limit
\begin{equation}
\theta_j \approx \frac{E_{\mathrm{jet}}-E_j}{E_{\mathrm{jet}}} 
\theta_{jk} \;\; .
\end{equation}
In the same way, one finds for the angle $\theta_k$ between 
parton $k$ and the direction of the jet
\begin{equation}
\theta_k \approx \frac{E_j}{E_{\mathrm{jet}}} \theta_{jk} \;\; .
\end{equation}
It is important to keep in mind that neither of the two particles $j$ and $k$ 
is allowed to be outside the cone. In other words, the upper limit
$\theta_{\mathrm{max}}$ in Eq.~(\ref{eq:scaps3}) in terms of the cone opening
$\delta$ is
\begin{eqnarray}
\label{eq:regions}
&& E_k >E_j\, : \;\;\; \theta_{\mathrm{max}}=\delta \; ,\nonumber \\
&& E_j >E_k\, : \;\;\; \theta_{\mathrm{max}}=\frac{E_k}{E_j}\delta
\;\; .
\end{eqnarray}
We note that the integration over $\theta_j$ in Eq.~(\ref{eq:scaps3}) 
could also be written as an integration over the invariant mass 
$\sqrt{P_{\mathrm{jet}}^2}$ of the jet. 
Consistently with the SCA, we have previously neglected the jet mass everywhere. 
However, the singularity of the cross section at $\theta_{jk}=0$, where 
the jet mass vanishes, requires us to integrate over a finite region
$0\leq\ P_{\mathrm{jet}}^2\leq {\cal O}(\delta^2 E_{\mathrm{jet}}^2)$ here. 

We finally obtain, after performing the $\theta_j$ integration
in Eq.~(\ref{eq:scaps3}) and introducing the variable 
$\xi=E_j/E_{\mathrm{jet}}$, 
\begin{eqnarray} 
\label{eq:scapsfinal}
\int \frac{dPS_3}{dvdw} \frac{1}{(2p_j\cdot p_k)} &=&
\left[ \frac{1}{8\pi} \left( \frac{4\pi}{s} 
\right)^{\varepsilon}\frac{1}{\Gamma(1-\varepsilon)}[v(1-v)]^{-\varepsilon}
\right] \frac{1}{8\pi^2} \left( \frac{4\pi}{s}
\right)^{\!\varepsilon} \delta(1-w)\nonumber \\[3mm]
&&\hspace*{-3.75cm}
\times\frac{1}{\Gamma(1-\varepsilon)}\left( -\frac{1}{\varepsilon} \right)
\left(\frac{E_{\mathrm{jet}}^2 \delta^2}{s}
\right)^{-\varepsilon}
\int_0^1 d\xi \left[ \xi^{-2 \varepsilon}\; \Theta\!
\left( \frac{1}{2}-\xi \right)+
(1-\xi)^{\!-2 \varepsilon}\; \Theta\!\left( \xi -\frac{1}{2}\right) 
\right] \;.
\end{eqnarray}
Here, $\Theta$ denotes the Heaviside step-function, delineating the 
two regions specified in Eq.~(\ref{eq:regions}).

Eq.~(\ref{eq:collapprox}) obviously continues to hold. 
The difference between the calculation of $d\Delta\hat{\sigma}_{j(k)}$ 
in Sec.~II.3 and of $d\Delta\hat{\sigma}_{jk}$ here is that we now need 
to integrate over the argument of the splitting function
appearing in Eq.~(\ref{eq:collapprox}), as $\xi$ in Eq.~(\ref{eq:scapsfinal}) 
plays the role of the momentum fraction $z$. To do so, let us denote
\begin{equation}
I_{mn}\equiv \int_0^1 d\xi \left[ \xi^{-2 \varepsilon}\;
\Theta\left( \frac{1}{2}-\xi \right)+
(1-\xi)^{-2 \varepsilon}\; \Theta\left( \xi -\frac{1}{2}\right) \right] 
P_{mn}^{<}(\xi) \;\; .
\end{equation}
Then one straightforwardly computes the four relevant integrals $I_{mn}$:
\begin{eqnarray}
I_{qq}&=& C_F \left[ -\frac{1}{\varepsilon} -\frac{3}{2} +\varepsilon
\left(  -\frac{7}{2}+\frac{\pi^2}{3} -3 \ln(2) \right) \right] \;=\;
I_{gq}\; , \nonumber \\
I_{qg}&=& \frac{1}{2} \left[\frac{2}{3} +\varepsilon
\left(  \frac{23}{18}+\frac{4}{3}\ln(2) \right) \right] \; , \nonumber \\
I_{gg}&=&2 C_A \left[ -\frac{1}{\varepsilon} -\frac{11}{6} +\varepsilon
\left( -\frac{137}{36} +\frac{\pi^2}{3} -\frac{11}{3} \ln(2) \right) \right]
\;\;,
\label{eq:imn}
\end{eqnarray}
with $C_A=3$ and $C_F=4/3$ the usual SU(3) Casimir operators.
Note that the $1/\varepsilon$ pole terms in Eqs.~(\ref{eq:imn}), along 
with the overall $1/\varepsilon$ singularity in Eq.~(\ref{eq:scapsfinal}), 
give double poles that exactly cancel the ones arising for collinear 
splittings in the $d\Delta\hat{\sigma}_{j(k)}$ (see the discussion at the end 
of Sec.~II.3). This cancelation will be demonstrated in the next section.
Furthermore,  the term $-3/2$ in the first line of Eqs.~(\ref{eq:imn})
is identical to the negative of the coefficient of the $\delta (1-z)$ in the 
full $P_{qq}$ splitting function.
It is extremely important that we obtain this term: clearly,
the factorization subtraction in Eq.~(\ref{eq:scafact}) contains the 
{\em full} splitting function, whereas all collinear limits
of $2\to 3$ matrix elements, Eq.~(\ref{eq:collapprox}), give only 
the splitting functions {\em without} their $\delta (1-z)$ contributions. 
Without the term $-3/2$ above we would not be able to cancel the final-state 
singularities. The same happens if the intermediate
parton $I$ is a gluon: an outgoing gluon may split into a $gg$ or a 
$q\bar{q}$ pair, the latter coming in $N_f$ ``active'' flavors.
Again we find in Eqs.~(\ref{eq:imn}) the necessary $-11/6$ and $2/3$ factors, 
which combine to the LO QCD $\beta$-function, $\beta_0=11C_A/3-2N_f/3$, 
appearing in the $\delta (1-z)$ contribution to $P_{gg}$.

%%%%%%%%%%%%%%%%%%%%%%%%%%%%%%%%%%%%%%%%%%%%%%%%%%%%%%%%%%%%%%%%
\subsection{Collecting Results and Cancelation of Singularities}
%%%%%%%%%%%%%%%%%%%%%%%%%%%%%%%%%%%%%%%%%%%%%%%%%%%%%%%%%%%%%%%%
%
The rest of the calculation is essentially an exercise in bookkeeping
of all of the many intermediate terms appearing in the various 
contributions to Eq.~(\ref{eq:jetoutline}).

To be more specific, let us sketch an example: a process with
a $q\bar{q}g$ final state. We already know $d\Delta\hat{\sigma}_q$,
$d\Delta\hat{\sigma}_{\bar{q}}$, and $d\Delta\hat{\sigma}_g$ from
the inclusive hadron calculation \cite{ref:nlopion}. The pieces
$d\Delta\hat{\sigma}_{g(q)}$ and $d\Delta\hat{\sigma}_{g(\bar{q})}$
rely on the splitting function $P_{gq}$ and are treated by 
Eq.~(\ref{eq:scanondiag}). Likewise, $d\Delta\hat{\sigma}_{q(\bar{q})}$
is related to $P_{qg}$. One only has to attach the appropriate LO
cross section [depending on $v'$ as defined above in 
Eq.~(\ref{eq:collapprox})]. On the other hand,
$d\Delta\hat{\sigma}_{q(g)}$ and $d\Delta\hat{\sigma}_{\bar{q}(g)}$ 
depend on the diagonal splitting function $P_{qq}$ and are singular.
It is therefore convenient to combine them directly with 
$d\Delta\hat{\sigma}_{qg}$ and $d\Delta\hat{\sigma}_{\bar{q}g}$, 
respectively. The combination  $d\Delta\hat{\sigma}_{q(g)} - 
d\Delta\hat{\sigma}_{qg}$ is, after
$\overline{\rm{MS}}$ subtraction of final-state collinear 
singularities at a scale $\mu^{\prime}_F$, proportional to
\begin{eqnarray}
\nonumber
d\Delta\hat{\sigma}_{q(g)} - d\Delta\hat{\sigma}_{qg} &\!\!\propto\!\!& 
\frac{\alpha_s}{2\pi}\,C_F\left\{ \delta (1-w)
\left[ \left( 2 \ln v +\frac{3}{2} \right)
\ln\! \left(\frac{\delta^2 E_{\mathrm{jet}}^2}{{\mu^{\prime}_F}^{\!2}} 
\right)+2 \ln^2 v
-\frac{7}{2} + \frac{\pi^2}{3}-3 \ln(2) \right] \right.\nonumber \\[3mm]
&\!\!+\!\!& \frac{2}{(1-w)_+} \left[ 2 \ln v+\ln \left(
\frac{\delta^2 E_{\mathrm{jet}}^2}{{\mu^{\prime}_F}^{\!2}} \right) \right]+
4 \left[ \frac{\ln(1-w)}{1-w} \right]_+\nonumber \\[3mm]
&\!\!+\!\!&\left.  \frac{v^2 (1-w)}{1-v+vw}\left[1
+\ln \left(\frac{\delta^2 E_{\mathrm{jet}}^2 v^2
    (1-w)^2}{{\mu^{\prime}_F}^{\!2}} 
\right) \right] \right\} \; .
\label{eq:qgblock}
\end{eqnarray}
This result is completely finite and may serve as a ``building
block'' whenever there is an outgoing quark radiating a gluon. 
Only the appropriate longitudinally polarized LO cross section 
$d\Delta\hat{\sigma}_{ab\rightarrow ql}/dv$ has to be attached to 
Eq.~(\ref{eq:qgblock}), summed over all final-state partons $l$.
The ``+''-distribution in Eq.~(\ref{eq:qgblock}) is defined 
as usual by its integration with an appropriate test function:
\begin{equation}
\int_0^1 dw \,\left[g(w)\right]_+ \, f(w) =
\int_0^1 dw \,g(w) \,\left[f(w)-f(1) \right] \; .
\end{equation}

The final piece in this example is $d\Delta\hat{\sigma}_{q\bar{q}}$ which 
involves the integral $I_{qg}$ in Eq.~(\ref{eq:imn}) and therefore is
also singular. It is crucial here that we sum over {\em all} 
possible final state channels for two given colliding partons:
if there is a contribution involving $d\Delta\hat{\sigma}_{q\bar{q}}$, 
there must also be one involving $d\Delta\hat{\sigma}_{gg}$.
With the appropriate combinatorial prefactors these two will
combine to a $\beta_0$, as discussed above. Including also 
$d\Delta\hat{\sigma}_{g(g)}$, and after $\overline{\rm{MS}}$ subtraction
of final-state collinear singularities, one obtains
for the ``building block'' for intermediate gluons, i.e.\ $I=g$,
\begin{eqnarray}
\nonumber
2 \,d\Delta\hat{\sigma}_{g(g)} - 2 N_f \, d\Delta\hat{\sigma}_{q\bar{q}}
-  d\Delta\hat{\sigma}_{gg} && \\[3mm]
&&\hspace*{-5.5cm} \propto \frac{\alpha_s}{2\pi}\left\{ \delta (1-w)
\left[ \left( 2 C_A \ln v +\frac{\beta_0}{2} \right)
\ln \left(\frac{\delta^2 E_{\mathrm{jet}}^2}{{\mu^{\prime}_F}^{\!2}} \right)+
C_A \left( 2\ln^2 v+\frac{\pi^2}{3}-\frac{17}{8}\right)\right.
\right.\nonumber \\[3mm]
&&\hspace*{-5.5cm} -\left.
\frac{11}{24}\beta_0+\frac{1}{3} N_f-\beta_0 \ln(2) \right] 
+ \frac{2C_A}{(1-w)_+} \left[ 2 \ln v+\ln \left(
\frac{\delta^2 E_{\mathrm{jet}}^2}{{\mu^{\prime}_F}^{\!2}} \right) \right]+
4 C_A\left[ \frac{\ln(1-w)}{1-w} \right]_+\nonumber \\[3mm]
&&\hspace*{-5.5cm} +\left. 2 C_A \frac{v^2 (1-w)}{(1-v+vw)^2}
\left[ 1+(1-v)^2 +v w (2-2v+vw)\right]
\ln \left(\frac{\delta^2 E_{\mathrm{jet}}^2 v^2 
(1-w)^2}{{\mu^{\prime}_F}^{\!2}} \right) 
\right\} \;,
\label{eq:gblock}
\end{eqnarray}
again finite. Recall that contributions from 
$d\Delta\hat{\sigma}_{q(\bar{q})}$, which also appear in 
Eq.~(\ref{eq:jetoutline}), are associated with the non-diagonal 
splitting function $P_{qg}$ and thus are finite by themselves and 
hence not included in (\ref{eq:gblock}). They are straightforwardly 
treated according to Eq.~(\ref{eq:scanondiag}). Our main equations,
Eqs.~(\ref{eq:scanondiag}), (\ref{eq:qgblock}), and (\ref{eq:gblock}),
all show the expected ``logarithmic plus constant'', i.e., 
${\cal A}\log(\delta)+{\cal B}+{\cal O}
(\delta^2)$, dependence on the jet cone opening $\delta$. 
We note that the terms $\propto \left[ \ln(1-w)/(1-w) \right]_+$ in 
Eqs.~(\ref{eq:qgblock}) and (\ref{eq:gblock}), which 
are the leading contributions at $w\to 1$, cancel identical 
pieces arising from the observed final-state parton in the single-parton 
inclusive cross sections $d\Delta\hat{\sigma}_q$, $d\Delta\hat{\sigma}_g$,
so that such distributions are absent in the jet cross section. This
finding is in accordance with results found in a large-$w$ ``threshold''
resummation calculation of the jet cross section, when the
jet is allowed to be massive at partonic threshold \cite{ref:KOS}.  

With these prerequisites at hand one can compute all relevant NLO
partonic single-inclusive jet cross sections 
listed in Eq.~(\ref{eq:proclist}) in the SCA.
The final analytical results are too lengthy to be given here but can be found
in a {\sc Fortran} code which is available upon request from the authors.
As mentioned already, a powerful check for the correctness of the
calculation is the cancelation of any dependence on the final state 
factorization scale $\mu^{\prime}_F$ when, according to
Eq.~(\ref{eq:jetoutline}), the single-parton-inclusive cross sections
from \cite{ref:nlopion} are combined with the appropriate combinations of
$d\Delta\hat{\sigma}_{j(k)}$ and $d\Delta\hat{\sigma}_{jk}$ according to 
Eqs.~(\ref{eq:scanondiag}), (\ref{eq:qgblock}), and
(\ref{eq:gblock}). All of our final results pass, of course, 
this consistency check. Another important check of the procedure 
outlined here comes from the computation of the unpolarized jet cross 
section in the SCA -- the building blocks in Eqs.~(\ref{eq:qgblock}) 
and (\ref{eq:gblock}) have no memory of the polarization of the 
colliding partons which only enters through the LO $2\to 2$ cross 
sections attached to them. We fully agree at an analytical
level with the results of \cite{ref:aversa} which can be retrieved 
from their {\sc Fortran} code, after an appropriate transformation 
to the $\overline{\mathrm{MS}}$ factorization scheme.

%%%%%%%%%%%%%%%%%%%%%%%%%%%
\section{Numerical Results}
%%%%%%%%%%%%%%%%%%%%%%%%%%%
%
%%%%%%%%%%%%
% FIGURE 3 %
%%%%%%%%%%%%
\begin{figure}[th]
\vspace*{-0.75cm}
\begin{center}
\begin{minipage}{8.7cm}
\begin{center}
\epsfig{figure=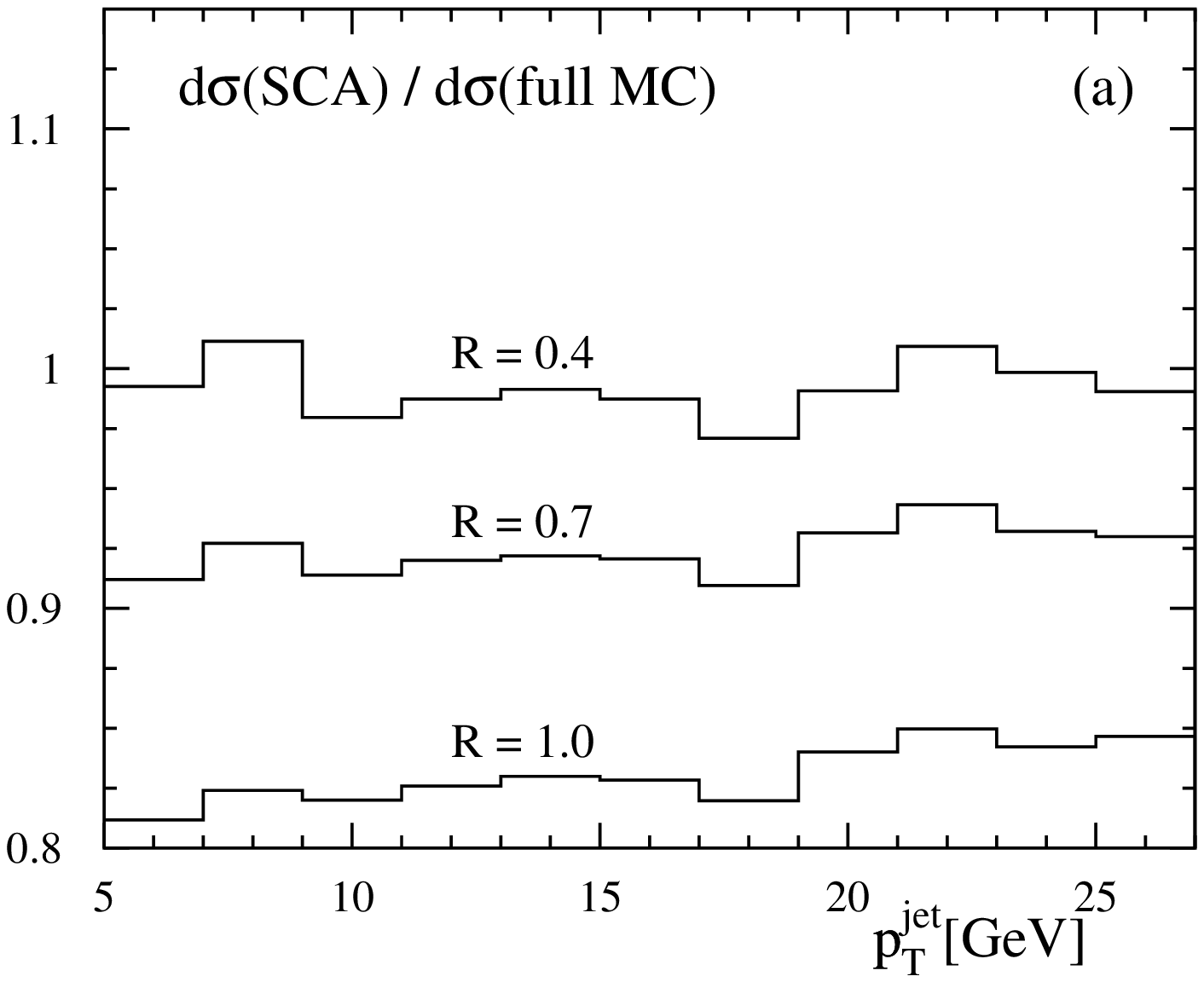,angle=0,width=1.05\textwidth}
\end{center}
\end{minipage}
\hspace*{-1.0cm}
\begin{minipage}{8.7cm}
\begin{center}
\epsfig{figure=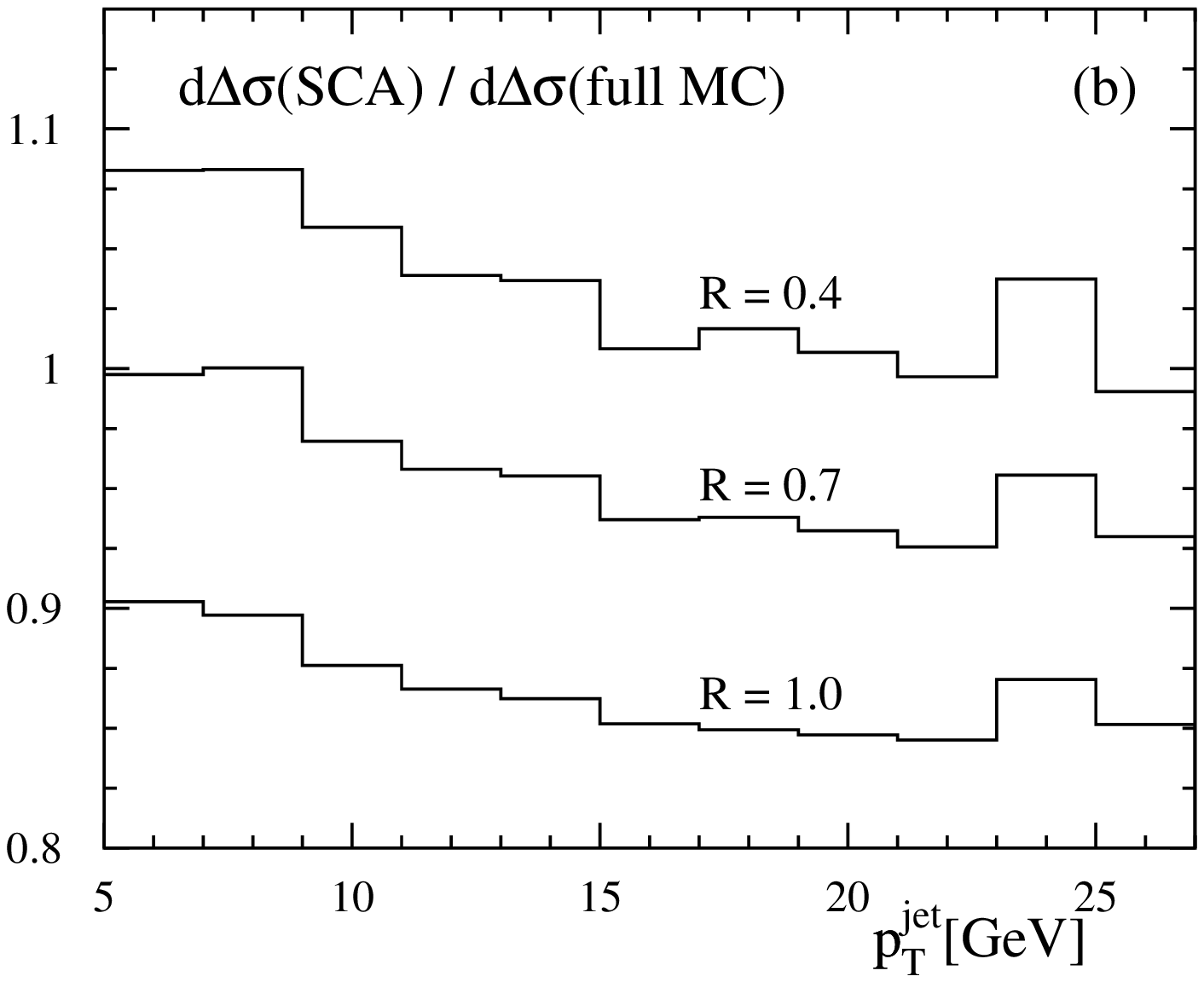,angle=0,width=1.05\textwidth}
\end{center}
\end{minipage}
\end{center}
\vspace*{-0.6cm}
\caption{\sf Ratio of the NLO {\bf (a)} unpolarized 
and {\bf (b)} polarized inclusive jet cross section 
in the SCA and within the full Monte-Carlo approach of \cite{ref:mcjets}
for $\sqrt{S}=200\,\mathrm{GeV}$, several bins in $p_T^{\mathrm{jet}}$, and 
three different cone sizes $R$. \label{fig:fig8}}
\end{figure}
We now turn to a phenomenological study of single-inclusive 
high-$p_T$ jet production in polarized $pp$ collisions at RHIC.
Regarding our results obtained in the SCA, the most important
question is, of course, how accurate the approximation actually is
in cases of practical relevance. We will investigate this first, 
by comparing our results to those of the Monte-Carlo code 
of \cite{ref:mcjets}. For this purpose, we choose
the recent CTEQ6M set of unpolarized parton densities \cite{ref:cteq6} 
in the calculation of the unpolarized cross section, and the NLO 
``standard'' set of GRSV \cite{ref:grsv} for the polarized case.
We assume the kinematical coverage of the {\sc Star} experiment, 
which can detect jets in the pseudo-rapidity range $-1\le 
\eta^{\mathrm{jet}} \le 1$, over which we integrate.

In Fig.~\ref{fig:fig8} we compare the results of the full Monte-Carlo 
NLO jet calculation \cite{ref:mcjets} to the results within the SCA 
for $\sqrt{S}=200\,\mathrm{GeV}$, several bins in $p_{T}^{\mathrm{jet}}$, 
and three different cone sizes $R$. 
The histograms on the left show the unpolarized results, the right ones give the 
comparison for the polarized case. It turns out that even for the 
rather large cone radius of $R=0.7$ the SCA gives still very acceptable results
within ten percent or less of the full Monte-Carlo calculation. 
For the unpolarized jet cross section such an observation was already made in
\cite{ref:scavalid,ref:guillet}. In the polarized case, however, the success of the
SCA up to large $R$ is non-trivial due to possible cancelations
between the two helicity configurations in Eq.~{(\ref{eq:xsecdef}).
Not unexpectedly, for very large $R=1.0$ the SCA starts to 
break down because neglected contributions proportional to 
${\cal{O}}(R^2)$ become relevant then. It is 
expected though that a cone size between 0.4 and 0.7 will be
%
%%%%%%%%%%%%
% FIGURE 4 %
%%%%%%%%%%%%
\begin{figure}[th]
\vspace*{-0.75cm}
\begin{center}
\begin{minipage}{8.7cm}
\begin{center}
\epsfig{figure=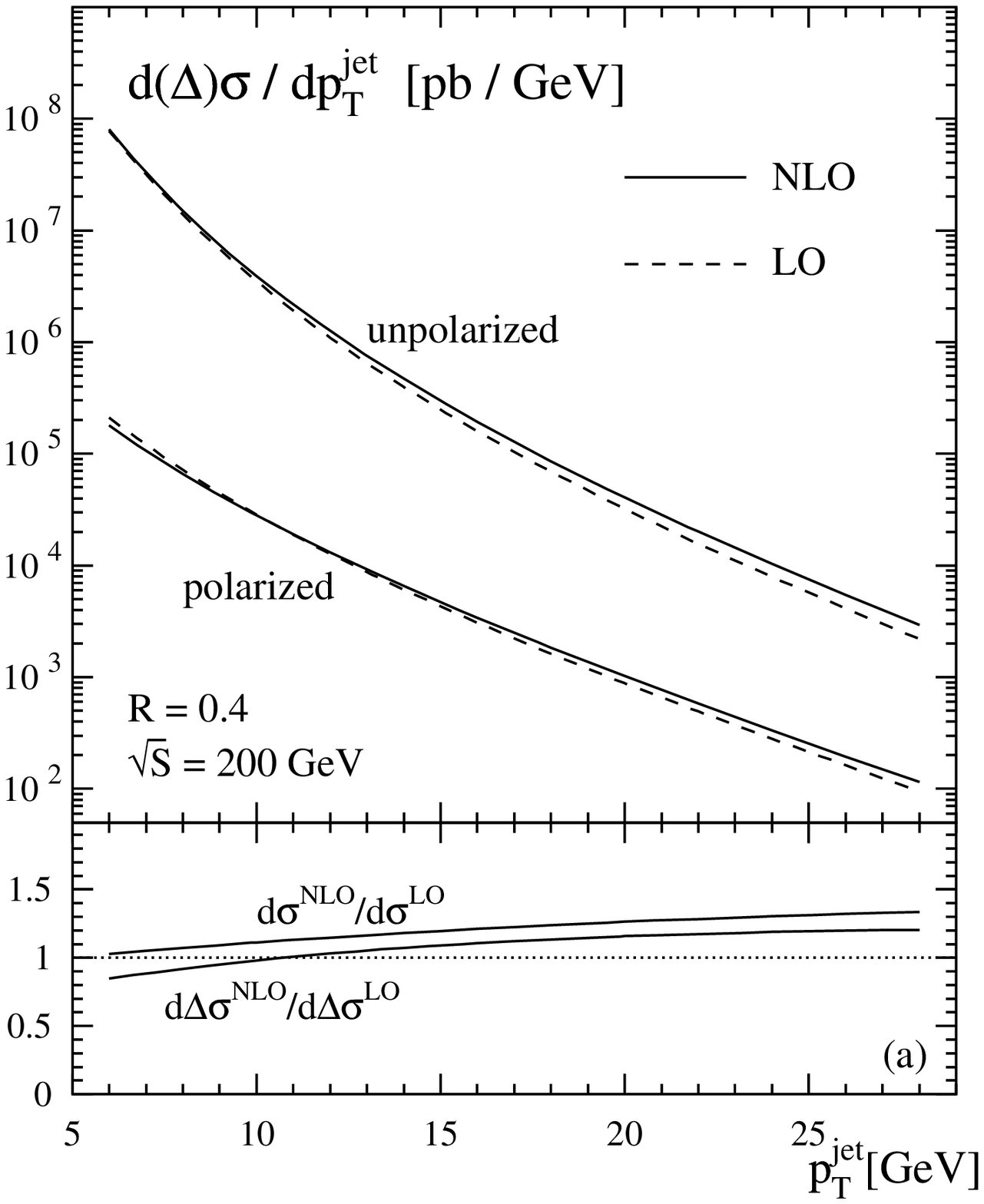,angle=0,width=0.99\textwidth}
\end{center}
\end{minipage}
\hspace*{-1.00cm}
\begin{minipage}{8.7cm}
\begin{center}
\epsfig{figure=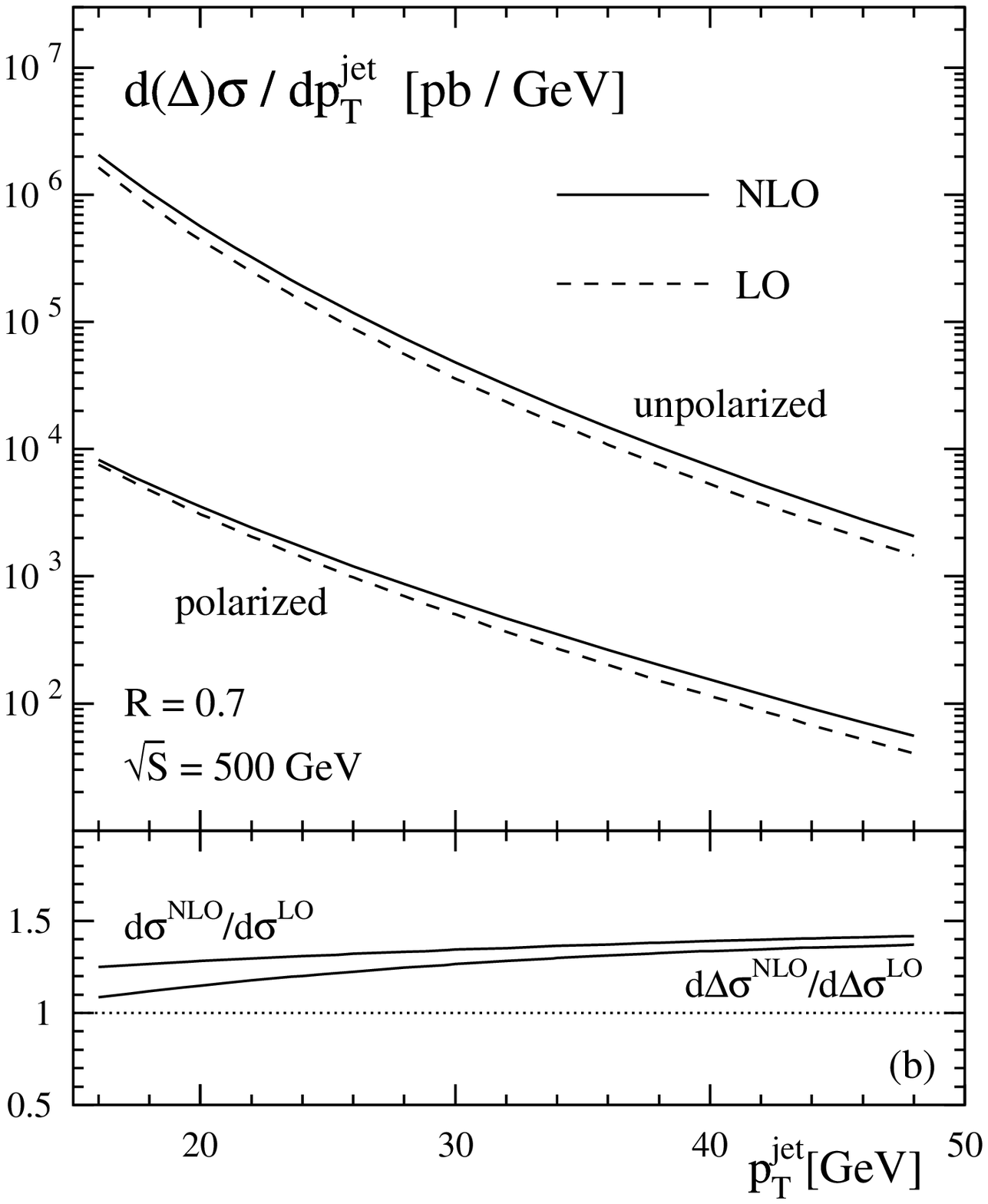,angle=0,width=0.99\textwidth}
\end{center}
\end{minipage}
\end{center}
\vspace*{-0.5cm}
\caption{\sf Unpolarized and polarized inclusive jet production cross sections
in NLO (solid) and LO (dashed lines) at {\bf (a)} $\sqrt{S}=200\,\mathrm{GeV}$ and
{\bf (b)} $\sqrt{S}=500\,\mathrm{GeV}$ for cone sizes $R=0.4$ and  $R=0.7$,
respectively. In each case the lower panel shows the ratios of the NLO and
LO results. \label{fig:fig3}}
\end{figure}
chosen by the {\sc Star} collaboration in their forthcoming analysis; 
larger sizes are not really practical in view of the limited angular 
acceptance of the detector. 

Fig.~\ref{fig:fig8} demonstrates that our results based on the
SCA are sufficiently accurate to be used in analyses of forthcoming 
data on jet cross sections and spin asymmetries from RHIC. We emphasize
that numerically stable results for the full $p_T^{\mathrm{jet}}$ spectrum 
can be obtained with our computer code very fast and efficiently, 
in a matter of minutes. This makes our code an ideal ingredient
for future ``global'' analyses of RHIC jet data in terms of 
polarized parton densities. This is a clear advantage over a Monte-Carlo 
code with its huge numerical complexity, which yields results with
rather large numerical fluctuations (still visible in the histograms 
shown in Fig.~\ref{fig:fig8}) 
even after hours of running. 
This is even more true for the polarized cross section
due to large cancelations between the two helicity configurations in
Eq.~(\ref{eq:xsecdef}).
That said, our present calculation can only be used to describe 
single-inclusive jet cross sections, whereas the Monte-Carlo code 
is much more flexible concerning the observables that can be predicted.

%%%%%%%%%%%%
% FIGURE 5 %
%%%%%%%%%%%%
\begin{figure}[th]
\vspace*{-0.75cm}
\begin{center}
\begin{minipage}{8.65cm}
\begin{center}
\epsfig{figure=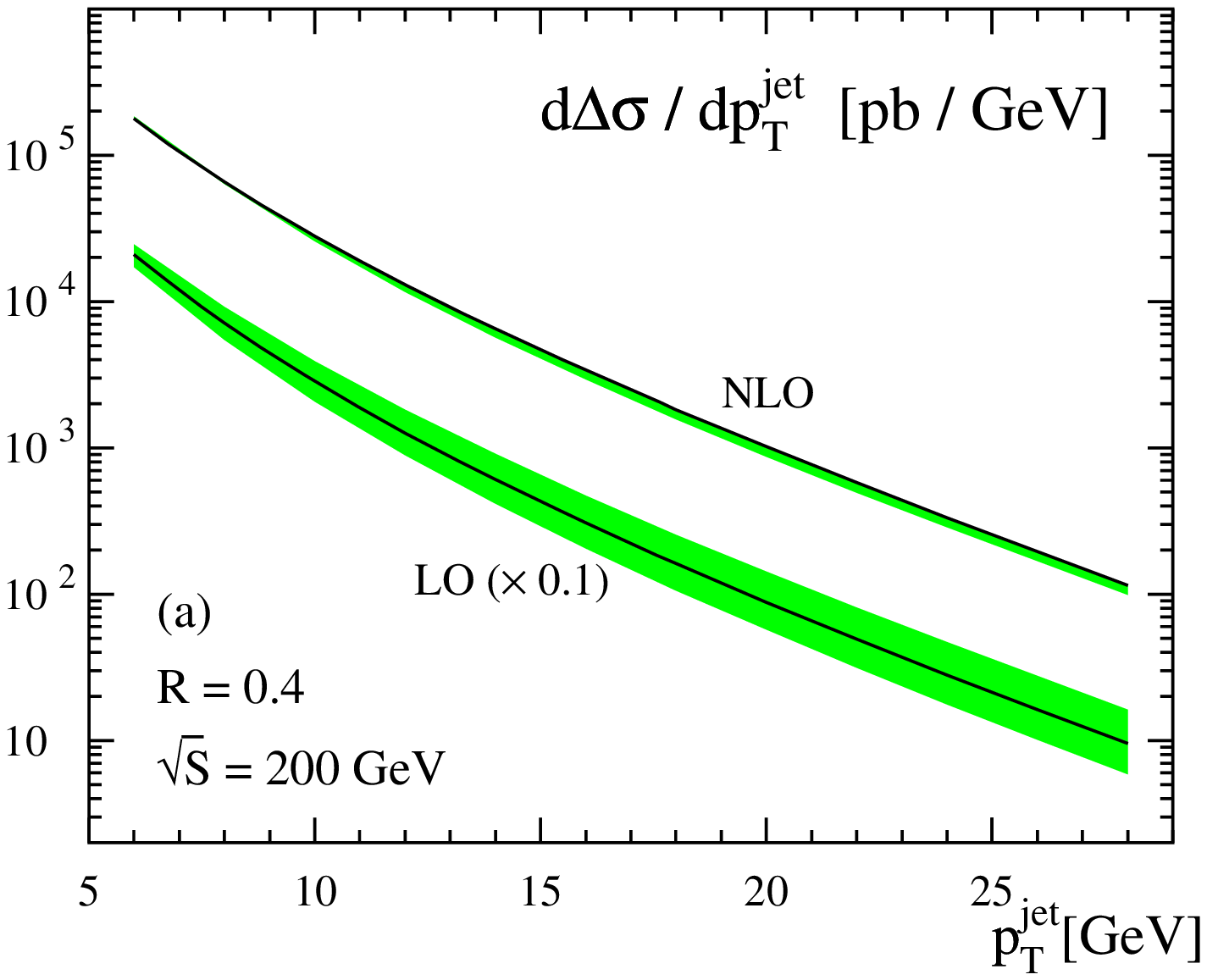,angle=0,width=1.05\textwidth}
\end{center}
\end{minipage}
\hspace*{-1.0cm}
\begin{minipage}{8.65cm}
\begin{center}
\epsfig{figure=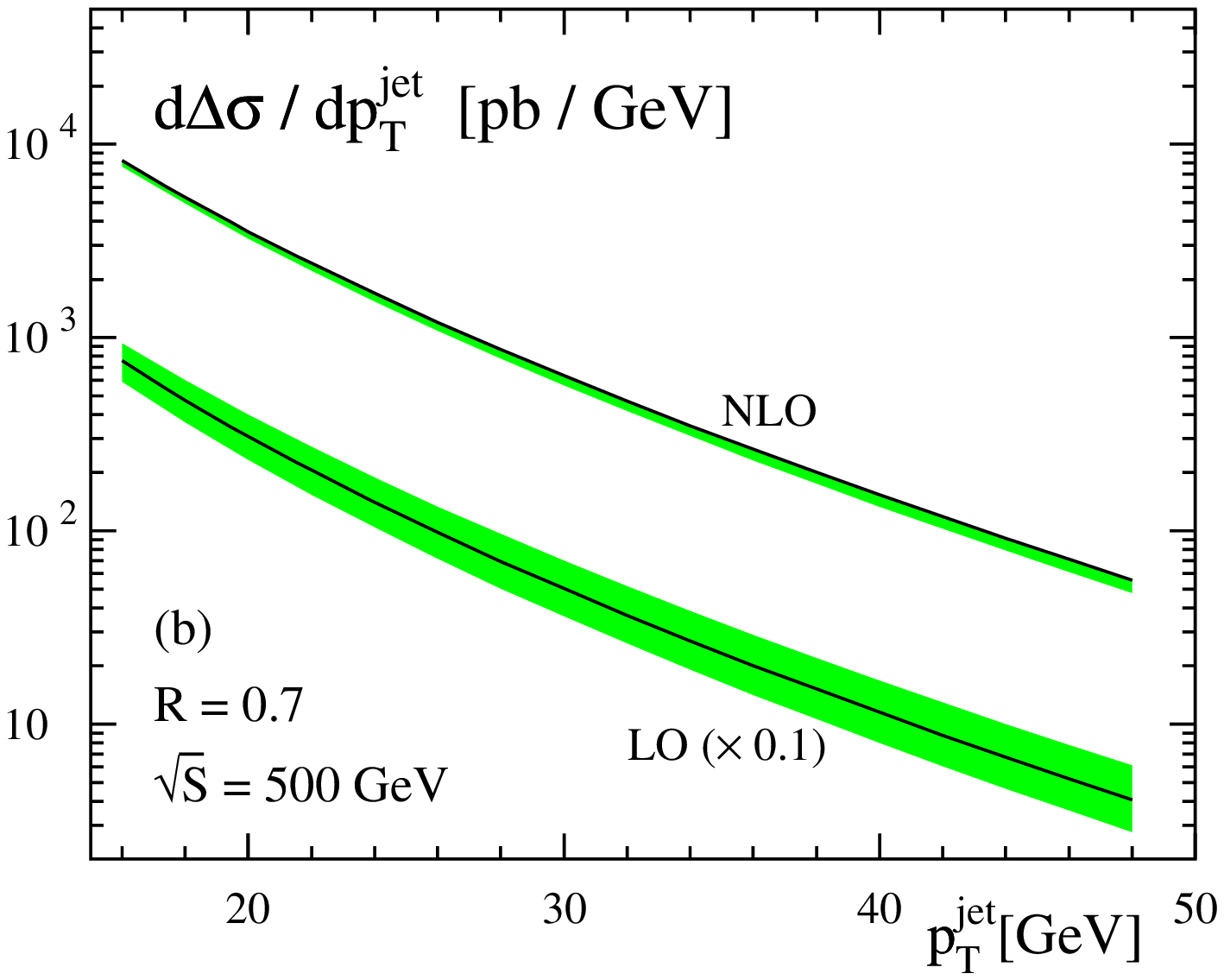,angle=0,width=1.05\textwidth}
\end{center}
\end{minipage}
\end{center}
\vspace*{-0.5cm}
\caption{\sf As in Figs.~\ref{fig:fig3} (a) and (b) but now showing the scale dependence 
of the polarized cross section in LO and NLO in the range
$p_T^{\mathrm{jet}}/2\le \mu_F = \mu_R \le 2 p_T^{\mathrm{jet}}$. 
We have rescaled the LO results by 0.1 to separate them better from the NLO ones. 
In each case the solid lines correspond to the choice where both scales are 
set to $p_T^{\mathrm{jet}}$.
\label{fig:fig4}}
\end{figure}
Heartened by the good agreement between our code and the 
full Monte-Carlo calculation, we will now present a few predictions for 
RHIC. We will be very brief here because many phenomenological 
results for jet production have already been presented in
\cite{ref:mcjets}. We focus on the most interesting questions:
the importance of the NLO corrections, the residual dependence on 
the unphysical scales $\mu_F$ and $\mu_R$ in Eq.~(\ref{eq:xsecfact}) at 
NLO, and the sensitivity of the double spin asymmetry 
\begin{equation}
\label{eq:all}
A_{\mathrm{LL}}^{\mathrm{jet}} \equiv \frac{d\Delta\sigma}{d\sigma}
\end{equation}
to the still unknown gluon polarization $\Delta g$.
Predictions for $A_{\mathrm{LL}}^{\mathrm{jet}}$ are in
immediate demand for an extraction of $\Delta g$ at RHIC in the 
very near future. In comparison to \cite{ref:mcjets}, there are
also some new results in our analysis.

Figure~\ref{fig:fig3} shows our results for the unpolarized and polarized
$p_T^{\mathrm{jet}}$-spectra of single-inclusive jets at NLO and LO, integrated over
$-1\le \eta^{\mathrm{jet}}\le 1$, for $\sqrt{S}=200\,\mathrm{GeV}$ and 
$\sqrt{S}=500\,\mathrm{GeV}$ and cone sizes $R=0.4$ and  $R=0.7$,
respectively. We have set the scales $\mu_R=\mu_F=p_T^{\mathrm{jet}}$. 
Again we have used CTEQ6M parton densities \cite{ref:cteq6} in the unpolarized
case and the GRSV ``standard'' set \cite{ref:grsv} for the polarized cross section,
always performing the NLO (LO) calculations of cross sections 
using NLO (LO) sets of parton distribution functions and the 
two-loop (one-loop) expression for $\alpha_s$. The lower part 
of the figure displays in each case the so-called ``$K$-factor''
\begin{equation}
\label{eq:kfactor}
K=\frac{d(\Delta)\sigma^{\rm NLO}}{d(\Delta)\sigma^{\rm LO}} \;\; .
\end{equation}
One can see that the NLO corrections are fairly moderate and of 
similar importance in the polarized and the unpolarized cases.

%%%%%%%%%%%%
% FIGURE 6 %
%%%%%%%%%%%%
\begin{figure}[th]
\vspace*{-0.75cm}
\begin{center}
\begin{minipage}{8.65cm}
\begin{center}
\epsfig{figure=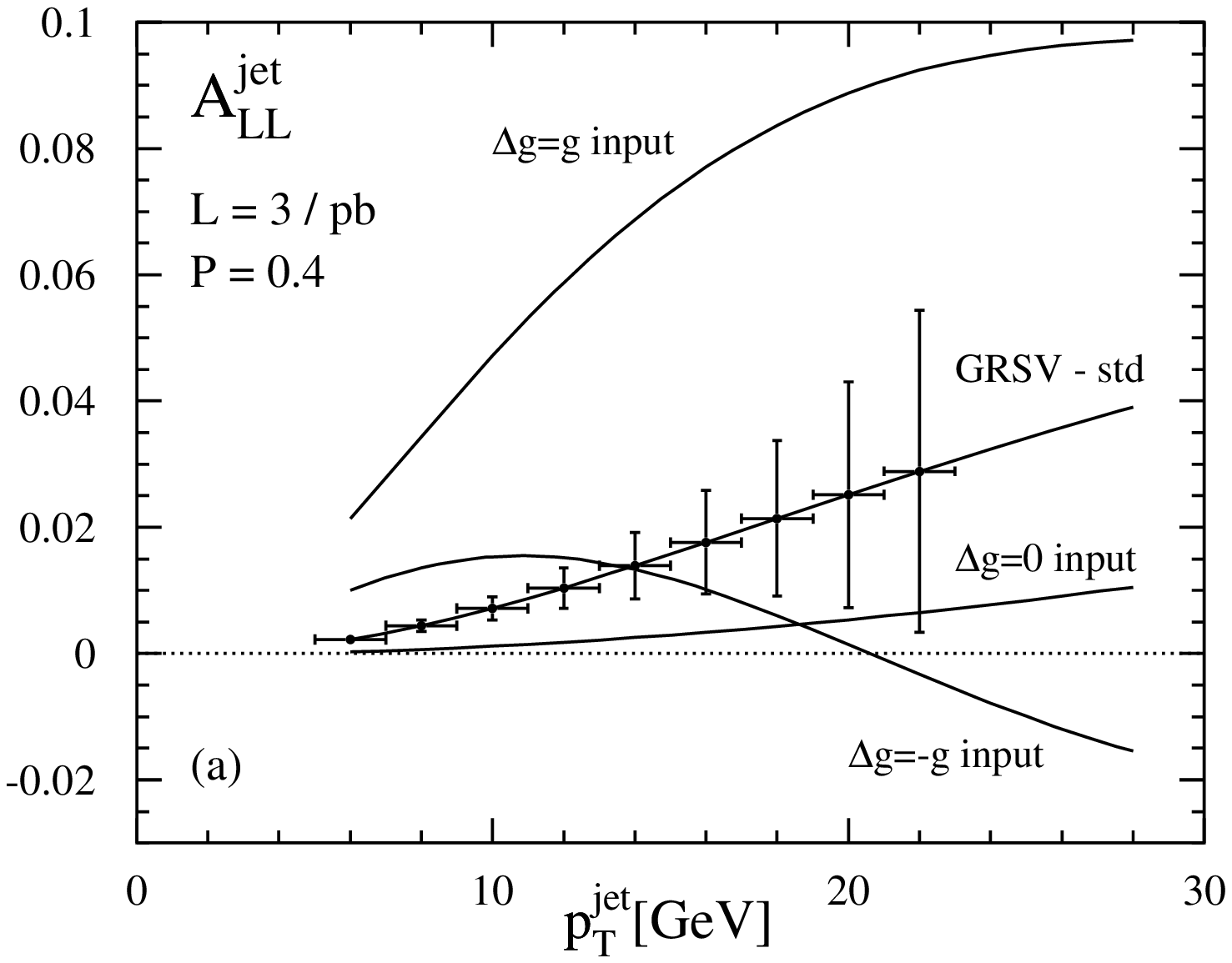,angle=0,width=1.05\textwidth}
\end{center}
\end{minipage}
\hspace*{-0.8cm}
\begin{minipage}{8.65cm}
\begin{center}
\epsfig{figure=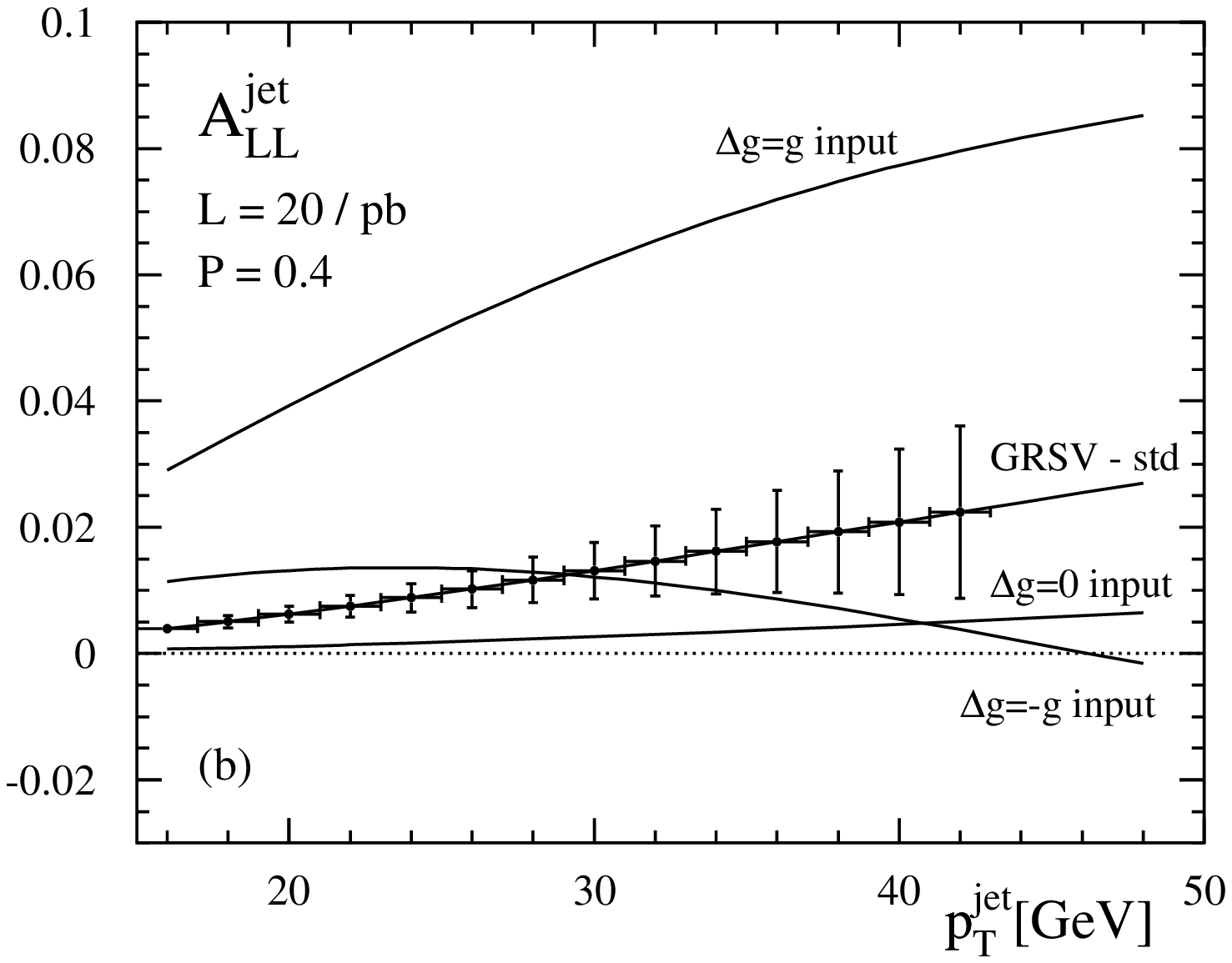,angle=0,width=1.05\textwidth}
\end{center}
\end{minipage}
\end{center}
\vspace*{-0.5cm}
\caption{\sf As in Figs.~\ref{fig:fig3} (a) and (b) but now showing the 
spin asymmetry at NLO using the GRSV ``standard'' set \cite{ref:grsv}
as well as three other sets with very different gluon polarizations (see text).
The ``error bars'' indicate the expected statistical accuracy 
$\delta A_{\mathrm{LL}}^{\mathrm{jet}}$ for $40\%$ beam polarization 
and integrated luminosities of $3\;\mathrm{pb}^{-1}$ and $20\;
\mathrm{pb}^{-1}$ for $\sqrt{S}=200\,\mathrm{GeV}$ and 
$\sqrt{S}=500\,\mathrm{GeV}$, respectively (see text).
\label{fig:fig5}}
\end{figure}
Figure~\ref{fig:fig4} shows the scale dependence of the spin-dependent
jet cross section at LO and NLO, again at $\sqrt{S}=200\,\mathrm{GeV}$ and 
$\sqrt{S}=500\,\mathrm{GeV}$ as in Figs.~\ref{fig:fig3} (a) and (b), respectively. 
In each case the shaded bands indicate the uncertainties from varying the
unphysical scales in the range $p_T^{\mathrm{jet}}/2 \le \mu_R=\mu_F\le 2 p_T^{\mathrm{jet}}$. 
The solid lines are for the choice where both scales are set to $p_T^{\mathrm{jet}}$.
Clearly, the scale dependence indeed becomes much smaller at NLO, 
a result that was already noted in \cite{ref:mcjets}. We emphasize 
that the scale ambiguity of the single-inclusive jet cross section
is somewhat smaller than for the corresponding single-inclusive hadron
cross section, cf., Fig.~5 in \cite{ref:nlopion}. This is not entirely
unexpected, as the additional final-state factorization scale $\mu^{\prime}_F$
related to the fragmentation of a parton into the observed hadron
provides a further source of scale dependence not present for jet production.

Next we consider the double spin asymmetry $A_{\mathrm{LL}}^{\mathrm{jet}}$, 
Eq.~(\ref{eq:all}), which is the main quantity of interest in experiment. 
Fig.~\ref{fig:fig5} shows $A_{\mathrm{LL}}^{\mathrm{jet}}$, calculated at
NLO for the ``standard'' set of GRSV parton densities \cite{ref:grsv}, and
for three other sets emerging from the GRSV analysis, which mainly differ 
in the assumptions about the gluon polarization: ``$\Delta g=g$ input'', 
``$\Delta g=0$ input'', and ``$\Delta g=-g$ input''. These are 
characterized by a large positive, a vanishing,
and a large negative gluon polarization, respectively, at the input scale
of the GRSV analysis \cite{ref:grsv}. We should note that all sets provide
a good description of all presently available data on spin-dependent 
deep-inelastic scattering. Again we show results for both c.m.s.\ energies 
relevant for RHIC; the other parameters are chosen as before.
Also shown is the expected statistical accuracy for such measurements in
certain bins of $p_T^{\mathrm{jet}}$ for the {\sc Star} experiment, calculated 
from \cite{ref:rhic}
\begin{equation}
\label{eq:allerror}
\delta A_{\mathrm{LL}}^{\mathrm{jet}} 
\simeq \frac{1}{{\cal{P}}_p^2 \sqrt{{\cal L}\sigma_{\rm bin}}} \; .
\end{equation}
%
%%%%%%%%%%%%
% FIGURE 7 %
%%%%%%%%%%%%
\begin{figure}[th]
\begin{center}
\epsfig{figure=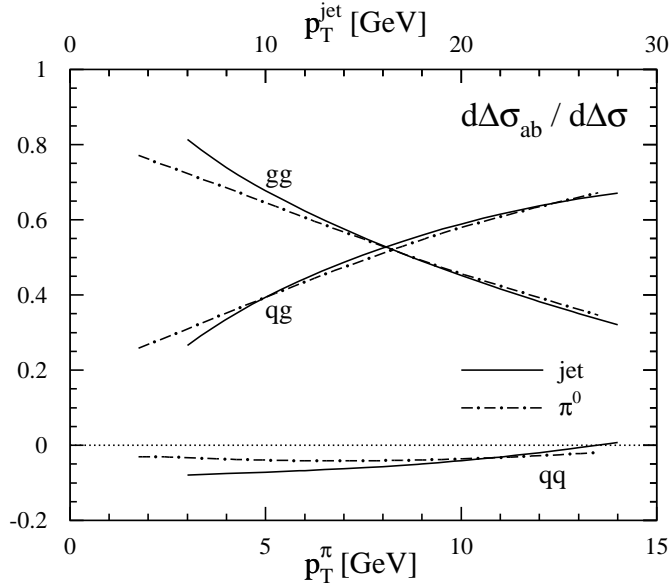,angle=0,width=0.60\textwidth}
\end{center}
\vspace*{-0.85cm}
\caption{\sf Relative contributions from $gg$, $qg$, and $qq$
scatterings to the NLO polarized cross section for the ``standard'' set 
of GRSV \cite{ref:grsv} for jet and inclusive-$\pi^0$ production at
mid pseudo-rapidities. Note
that we use two separate coordinate axes for the two cases.
\label{fig:fig7}}
\end{figure}
We assume a moderate proton beam polarization ${\cal{P}}_p$
of $40\%$. For the c.m.s.\ energy $\sqrt{S}=200\,\mathrm{GeV}$ 
we use an integrated luminosity ${\cal{L}}$ of only $3\,\mathrm{pb}^{-1}$ 
which should be well accomplishable in the near-term 
future. Also, we assume that only half of the calorimeter is instrumented, 
i.e., we integrate only over $0\le\eta^{\mathrm{jet}}\le 1$.
The error bars at $\sqrt{S}=500\,\mathrm{GeV}$ refer to 
${\cal{L}}=20\,\mathrm{pb}^{-1}$ and  $-1\le\eta^{\mathrm{jet}}\le 1$.
Statistical accuracies for higher beam polarizations and/or luminosities
are easily obtained by a proper rescaling of the error bars according to 
Eq.~(\ref{eq:allerror}).

First and foremost we conclude from Figs.~\ref{fig:fig5} that there are
excellent overall prospects for determining $\Delta g$ from 
$A_{\mathrm{LL}}^{\mathrm{jet}}$ measurements at RHIC: the spin asymmetries 
for the different sets of polarized parton densities, which mainly 
differ in the gluon density, show marked differences, much larger than 
the expected statistical errors in the experiment even for the very moderate
luminosities and beam polarizations assumed here. This finding was 
already highlighted in \cite{ref:mcjets}. It is interesting, however, 
that at moderately small $p_T^{\mathrm{jet}}$ the spin asymmetry 
$A_{\mathrm{LL}}^{\mathrm{jet}}$ is insensitive 
to the sign of $\Delta g$. A large negative gluon polarization 
yields an asymmetry somewhere in between the ones obtained
with moderately and large positive gluon polarizations. 
This observation is very similar to the one recently made for 
inclusive hadron production at small transverse momentum
\cite{ref:ourprl}. It is due to the fact that at moderate $p_T$ and
mid pseudo-rapidities the gluon-gluon initiated subprocess is dominant.
%
%%%%%%%%%%%%
% FIGURE 8 %
%%%%%%%%%%%%
\begin{figure}[th]
\vspace*{-0.75cm}
\begin{center}
\epsfig{figure=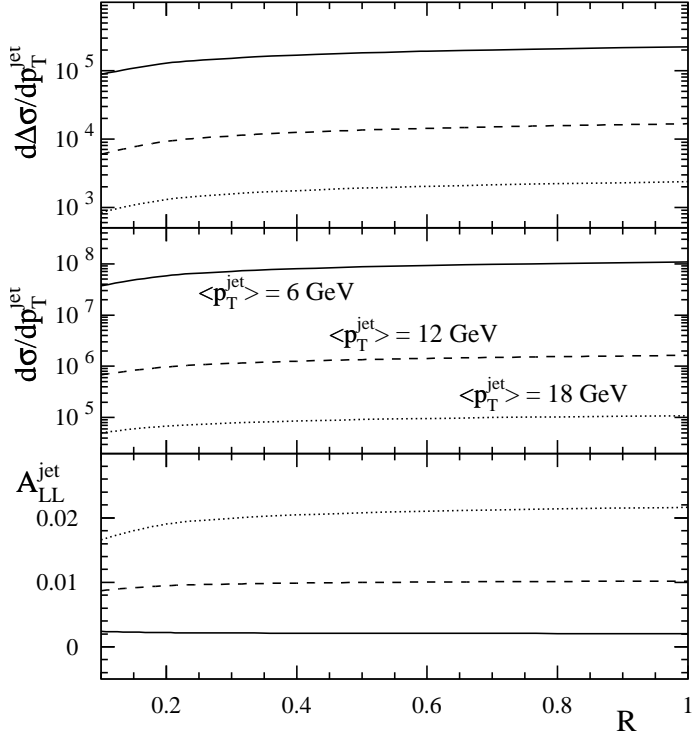,width=0.60\textwidth}
\end{center}
\vspace*{-0.8cm}
\caption{\sf Cone radius $R$ dependence of the polarized
and unpolarized single-inclusive jet cross section and the corresponding
spin asymmetry at NLO for $\sqrt{S}=200\,\mathrm{GeV}$ and 
three different $p_T^{\mathrm{jet}}$ bins. We have 
integrated over $-1\le\eta^{\mathrm{jet}}\le 1$. 
\label{fig:fig9}}
\end{figure}
This process has a positive analyzing power and, since the two gluons are
predominantly probed at very similar momentum fractions,
it essentially probes the square of $\Delta g(x_a\simeq x_b)$. Thus a 
positive asymmetry is obtained even for a negative $\Delta g$. At
larger jet transverse momenta, the $qg$ process gradually takes
over, resulting in sensitivity of $A_{\mathrm{LL}}^{\mathrm{jet}}$ to
the sign of $\Delta g(x_a\simeq x_b)$. When the c.m.s.~energy is 
higher, the onset of the $qg$ dominance occurs at higher
$p_T^{\mathrm{jet}}$ since the spin asymmetry roughly scales
with $x_T^{\mathrm{jet}}=2p_T^{\mathrm{jet}}/\sqrt{S}$.

These properties of the partonic scatterings are exemplified by 
the solid lines in Fig.~\ref{fig:fig7}, which show the relative 
contributions to the polarized NLO cross section for the GRSV 
``standard'' set for $gg$, $qg$, and $qq$ scatterings at 
$\sqrt{S}=200\,\mathrm{GeV}$. Here a 
``$q$'' stands generically for the appropriate sum of all quark and 
anti-quark flavors.
Each curve has been normalized to the full NLO cross section, so 
that the three lines add up to unity at every $p_T^{\mathrm{jet}}$.
For illustration, we also show in Fig.~\ref{fig:fig7} the 
corresponding results for inclusive $\pi^0$ production \cite{ref:nlopion}
at mid pseudo-rapidities.
We see that all curves for jets and pions are almost congruent, 
{\it provided} we rescale the axis for $p_T^{\pi}$ by about 
a factor of 2. This feature is understood from the fact
that pions result from a fragmentation process, in which the
pion inherits only a certain momentum fraction $z$ of a final-state
parton. At RHIC energies, mid pseudo-rapidities, and for the transverse momenta we are
considering here, one finds that the average $z$ is about 0.5. 
This means that a pion of, say $p_T^{\pi}=5$ GeV on average 
originates from a scattering in which a $10$ GeV parton was
produced. For the jet cross section, this parton would produce
however a jet with $p_T^{\mathrm{jet}}=10$ GeV. This explains
the results in Fig.~\ref{fig:fig7}. Similar relations are also 
found for the spin asymmetries $A_{\mathrm{LL}}^{\mathrm{jet}}$
and $A_{\mathrm{LL}}^{\pi}$. This interesting predicted interplay 
between hadron and jet observables may be exploited to cross-check 
results and to gain a deeper insight into the dynamics producing
high-transverse momentum final states. 

We close by showing in Fig.~\ref{fig:fig9} the dependence of the
polarized and unpolarized single-inclusive jet cross sections and the
corresponding spin asymmetry $A_{\mathrm{LL}}^{\mathrm{jet}}$ 
on the cone size $R$ in NLO QCD, for three representative 
bins in the jet's transverse momentum $p_T^{\mathrm{jet}}$. 
We have chosen $\sqrt{S}=200\,\mathrm{GeV}$ and again
integrated over pseudo-rapidity, $-1\le\eta^{\mathrm{jet}}\le 1$. The results for
$\sqrt{S}=500\,\mathrm{GeV}$ are very similar and not shown here.
We recall from Sec.~II that for $R$ not too large, the dependence 
on $R$ is logarithmic. Obviously, at least in the unpolarized case, 
the cross section has to rise with increasing $R$. At larger
$R$, deviations from our curves would be expected due to the 
terms $\propto R^2$ becoming important. A striking finding 
in Fig.~\ref{fig:fig9} is that $A_{\mathrm{LL}}^{\mathrm{jet}}$ 
turns out to depend only very mildly on $R$, in particular for 
$p_T^{\mathrm{jet}}$ not too large. We also note that since the 
cross section in Born approximation is independent of $R$, a measurement 
of the cone size dependence is a direct probe of NLO contributions.

%%%%%%%%%%%%%%%%%%%%%%%%%%%%%%%%%
\section{Summary and Conclusions}
%%%%%%%%%%%%%%%%%%%%%%%%%%%%%%%%%
%
In this paper we have presented a NLO calculation for the spin-dependent 
hadroproduction of single-inclusive jets. The application of the 
small cone approximation not only allowed us
to perform the calculation at a largely analytical level 
but also to use major parts of a previous calculation for
single-inclusive hadron production. We have outlined in some detail
the connection between these two cross sections. By comparing to
the Monte-Carlo jet code of \cite{ref:mcjets} which treats
the cone size exactly, we have demonstrated 
the applicability of the SCA up to cone sizes of about 0.7. Our
code has the advantage of being numerically very stable and fast.

Our results are useful for studies of large-$p_T$ jet production at
RHIC, in particular for the analysis of upcoming data in terms of the
unknown spin-dependent gluon density in the nucleon. We found that
the NLO corrections to the (un)polarized cross section are well under
control, and that their inclusion leads to a significant reduction of
scale dependence, the main source of theoretical ambiguity.
Even for rather moderate beam polarizations and integrated luminosities,
the corresponding spin asymmetry shows a sensitivity to $\Delta g$.
This makes single-inclusive jet production an excellent tool for fulfilling
the short-term goal of RHIC: a first determination of the gluon
polarization. Also in the long-run jet production data will provide
invaluable information in a detailed mapping of the $x$-shape of
spin-dependent parton densities.
The experimentally relevant spin asymmetry
turned out to be rather insensitive to the precise definition of the
jet as well as to the actual choice of the cone opening.

%%%%%%%%%%%%%%%%%%%%%%%%%%
\section*{Acknowledgments}
%%%%%%%%%%%%%%%%%%%%%%%%%%
%
We are grateful to D.\ de Florian, E.\ Laenen, and G.\ Sterman for very 
valuable discussions and to {J.-Ph.}~Guillet for providing us with 
matrix elements computed in \cite{ref:aversa}. 
B.J.\ and M.S.\ thank the RIKEN-BNL Research Center and 
Brookhaven National Laboratory for hospitality and support 
during different stages of this work and A.\ Sch\"{a}fer for discussions. 
W.V.\ is grateful to RIKEN, Brookhaven National Laboratory and the U.S.\
Department of Energy (contract number DE-AC02-98CH10886) for
providing the facilities essential for the completion of this work.
This work was supported in part by the 
``Bundesministerium f\"{u}r Bildung und Forschung (BMBF)''.

%%%%%%%%%%%%%%%%%%%%%%%%%%%

%

\begin{thebibliography}{99}
%%%%%%%%%%%%%%%%%%%%%%%%%%%
%
\bibitem{ref:rhic} See, for example: G.\ Bunce, N.\ Saito, J.\ Soffer, and 
W.\ Vogelsang, Annu.\ Rev.\ Nucl.\ Part.\ Sci.\ {\bf 50}, 525 (2000).
%
\bibitem{ref:phenixpion} {\sc Phenix} Collaboration, S.S.\ Adler 
{\em et al.}, Phys. Rev. Lett. {\bf 91}, 241803 (2003).
%
\bibitem{ref:starpion} {\sc Star} Collaboration, J.\ Adams {\em et al.},
{\tt hep-ex/0310058} (to appear in Phys. Rev. Lett.).
%
\bibitem{ref:phenixall} A.\ Bazilevsky, talk presented at the 
``X$^{\mathrm{th}}$
Workshop on High Energy Spin Physics (Spin-03)'', Dubna, Russia,
Sep.~16-20, 2003.
%
\bibitem{ref:ourprl} B.\ J\"{a}ger, M.\ Stratmann, S.\ Kretzer, 
and W.\ Vogelsang, Phys. Rev. Lett. {\bf 92}, 121803 (2004).
%
\bibitem{ref:nlopion} B.\ J\"{a}ger, A.\ Sch\"{a}fer, M.\ Stratmann,
and W.\ Vogelsang, Phys. Rev. {\bf D67}, 054005 (2003).
%
\bibitem{ref:daniel} D.\ de Florian, Phys. Rev. {\bf D67}, 054004 (2003).
%
\bibitem{ref:global} M.\ Stratmann and W.\ Vogelsang,
Phys. Rev. {\bf D64}, 114007 (2001).
%
\bibitem{ref:mcjets}D.\ de Florian, S.\ Frixione, A.\ Signer, 
and W.\ Vogelsang, Nucl. Phys. {\bf B539}, 455 (1999).
%
\bibitem{ref:sterwbg} G.\ Sterman and S.\ Weinberg, Phys. Rev. Lett.
{\bf 39}, 1436 (1977). 
%
\bibitem{ref:furman} M.A.\ Furman, Nucl. Phys. {\bf B197}, 413 (1982).
%
\bibitem{ref:aversa} F.\ Aversa, P.\ Chiappetta, M.\ Greco, and 
J.-Ph.\ Guillet, Nucl. Phys. {\bf B327}, 105 (1989);
Z.~Phys. {\bf C46}, 253 (1990).
%
\bibitem{ref:eks} S.D.\ Ellis, Z.\ Kunszt, and D.E.\ Soper,
Phys. Rev. Lett. {\bf 62}, 726 (1989); Phys. Rev. {\bf D40}, 2188 (1989);
Phys. Rev. Lett. {\bf 64}, 2121 (1990).
%
\bibitem{ref:snowmass} John E.\ Huth {\em et al.}, published in the proceedings of the 
1990 DPF ``Summer Study on High Energy Physics: Research Directions for the Decade'', Snowmass,
CO, USA, ed.\  E.L.\ Berger, World Scientific, 1992, p.~134.
%
\bibitem{ref:othercone} UA2 Collaboration, 
J.\ Alitti {\em et al.}, Phys. Lett. {\bf B257}, 232 (1991);\\
CDF Collaboration, F.\ Abe {\em et al.}, Phys. Rev. {\bf D45}, 1448 (1992);\\
D0 Collaboration, S. Abachi {\em et al.} Phys. Rev. {\bf D53}, 6000 (1996).
%
\bibitem{ref:cluster} S.\ Catani, Yu.L.\ Dokshitzer, M.H.\ Seymour, and B.R.\ Webber, 
Nucl. Phys. {\bf B406}, 187 (1993);\\
S.D.\ Ellis and D.E.\ Soper, Phys. Rev. {\bf D48} (1993) 3160.
%
\bibitem{ref:discussion} See also the discussions about jet definitions and algorithms in, e.g.:\\
W.B.\ Kilgore and W.T.\ Giele, Phys. Rev. {\bf D55}, 7183 (1997);\\
M.H.\ Seymour, Nucl. Phys. {\bf B513}, 269 (1998); proceedings of the
``8th International Workshop on Deep Inelastic Scattering and QCD (DIS 2000)'', 
Liverpool, England, 2000,
eds.\ J.A.\ Gracey and T.\ Greenshaw, World Scientific, 2001, p.\ 27.
%
\bibitem{ref:salesch} S.G.\ Salesch, Ph.D. thesis, Hamburg University, 1993, DESY-93-196
(unpublished).
%
\bibitem{ref:scavalid}  F.\ Aversa, P.\ Chiappetta, M.\ Greco, and 
J.-Ph.\ Guillet, Phys. Rev. Lett. {\bf 65}, 401 (1990);\\
 F.\ Aversa, P.\ Chiappetta, L.\ Gonzales, M.\ Greco, and 
J.-Ph.\ Guillet, Z. Phys. {\bf C49}, 459 (1991).
%
\bibitem{ref:guillet} J.-Ph.\ Guillet, Z. Phys. {\bf C51}, 587 (1991).
%
\bibitem{ref:fact} S.B.\ Libby and G.\ Sterman, Phys.\ Rev.\ {\bf D18}, 
3252 (1978); \\ R.K.\ Ellis, H.\ Georgi, M.\ Machacek, H.D.\ Politzer, 
and G.G.\ Ross, Phys. Lett. {\bf 78B}, 281 (1978); Nucl. Phys. {\bf B152}, 
285 (1979); \\ D.\ Amati, R.\ Petronzio, and G.\ Veneziano,
Nucl. Phys. {\bf B140}, 54 (1980); Nucl. Phys. {\bf B146}, 29 (1978);\\
G.\ Curci, W.\ Furmanski, and R.\ Petronzio, Nucl.\ Phys.\ {\bf B175}, 
27 (1980);\\ J.C.\ Collins, D.E.\ Soper, and G.\ Sterman, Phys.\ Lett.\ 
{\bf B134}, 263 (1984); Nucl.\ Phys.\ {\bf B261}, 104 (1985);\\
J.C.\ Collins, Nucl.\ Phys.\ {\bf B394}, 169 (1993).
%
\bibitem{ref:photcone} L.E.\ Gordon and W.\ Vogelsang, Phys. Rev. {\bf D50}, 1901 (1994).
%
\bibitem{ref:KOS} N.\ Kidonakis, G.\ Oderda, and G.\ Sterman,
Nucl. Phys. {\bf B525}, 299 (1998).
%
\bibitem{ref:cteq6} J.\ Pumplin {\em et al.}, JHEP {\bf 0207}, 012 (2002).
%
\bibitem{ref:grsv}  M.\ Gl\"{u}ck, E.\ Reya, M.\ Stratmann, and
W.\ Vogelsang, Phys. Rev. {\bf D63}, 094005 (2001). 
%
%%%%%%%%%%%%%%%%%%%%%%%%%%%%%%%%%%%%%%%%%%%%%%%%%%%%%5
%
\end{thebibliography}
\end{document}